\documentclass[aps,pra,twocolumn,a4paper,superscriptaddress,longbibliography,nofootinbib,notitlepage]{revtex4-1}
\usepackage{graphicx}
\usepackage{amsmath}
\usepackage{amssymb}

\usepackage{color}

\usepackage[english]{babel}
\usepackage[T1]{fontenc}
\usepackage{mathrsfs}
\usepackage[colorlinks]{hyperref}

\newcommand{\pton}[1]{\left(#1\right)}

\newcommand{\comm}[2]{\left[#1,#2\right]}
\newcommand{\der}[2]{\frac{\partial #1}{\partial #2}}

\newcommand{\md}[1]{\left|#1\right|}
\newcommand{\bk}[1]{\textbf{#1}}
\newcommand{\ave}[1]{\langle #1 \rangle}

\begin{document}

\title{Non-Markovian polaron dynamics in a trapped Bose-Einstein condensate}

\author{Aniello Lampo}\email{aniello.lampo@icfo.eu}
        \affiliation{ICFO -- Institut de Ciencies Fotoniques, The Barcelona Institute of Science and Technology, 08860 Castelldefels (Barcelona), Spain}
\author{Christos Charalambous}
        \affiliation{ICFO -- Institut de Ciencies Fotoniques, The Barcelona Institute of Science and Technology, 08860 Castelldefels (Barcelona), Spain}

\author{Miguel \'{A}ngel Garc\'{i}a-March}
        \affiliation{ICFO -- Institut de Ciencies Fotoniques, The Barcelona Institute of Science and Technology, 08860 Castelldefels (Barcelona), Spain}
\author{Maciej Lewenstein}
   \affiliation{ICFO -- Institut de Ciencies Fotoniques, The Barcelona Institute of Science and Technology, 08860 Castelldefels (Barcelona), Spain}

        \affiliation{ICREA -- Instituci{\'o} Catalana de Recerca i Estudis Avan\c{c}ats, Lluis Companys 23, E-08010 Barcelona, Spain}


\begin{abstract}
We study the dynamics of an impurity embedded in a trapped Bose-Einstein condensate, \textit{i.e.} the Bose polaron problem.
  This problem is treated by recalling open quantum systems techniques: the impurity corresponds to a particle performing quantum Brownian motion, while the excitation modes of the gas play the role of the environment.  It is crucial that the model considers a  parabolic trapping  potential to resemble the experimental conditions. Thus,  
  we detail here how the formal derivation changes due to the gas trap, in comparison to the homogeneous gas. More importantly, we elucidate all aspects in which the gas trap plays a relevant role, with an emphasis in the enhancement of the non-Markovian character of the dynamics. We first find that the presence of a gas trap leads to a new form of the bath-impurity coupling constant and a larger degree in the super-ohmicity of the spectral density. We then  solve the quantum Langevin equation to derive the position and momentum variances of the impurity, where the former is a measurable quantity.   
For the particular case of an untrapped impurity, the asymptotic behaviour of this quantity is found to be motion
super-diffusive.  
When the impurity is trapped, we find position squeezing, casting the system suitable for implementing quantum metrology and sensing protocols.
We detail how both super-diffusion and squeezing can be enhanced or inhibited by tuning the Bose-Einstein condensate trap frequency.  
Compared to the homogeneous gas case, the form of the bath-impurity coupling constant changes, and this is manifested as a different dependence of the system dynamics on the past history. To quantify this, we introduce several techniques to compare the different amount of memory effects arising in the homogeneous and inhomogeneous gas.
We find that it is higher in the second case.  This analysis paves the way to the study of non-Markovianity in ultracold gases, and the possibility to exploit such a property in the realization of new quantum devices. 

\end{abstract}
\date{\today}

\maketitle
\section{Introduction}
Quantum gases have sparked off theoretical and experimental scientific interest in recent years. 
They are an excellent testbed for many-body theory, and are particularly useful to investigate strongly coupled and correlated regimes, offering thus an interesting, sometimes even hard to reach alternative to condensed matter systems \cite{Bloch2008, Lewenstein2012}. 
The current work concerns the physics of an impurity in a Bose-Einstein condensate (BEC), intensively studied in the context of polaron physics in strongly-interacting Fermi~\cite{Schirotzek2009, Kohstall2012, Koschorreck2012,
MassignanPolRev2014, Lan2014, Levinsen2014, Schmidt2012} or Bose gases~\cite{Cote2002, Massignan2005, Cucchietti2006, Palzer2009, Catani2012, Cugliandolo2012,
Spethmann2012, Rath2013, Fukuhara2013, Bonart2013,
Shashi2014, Benjamin2014, Grusdt2014a, Grusdt2014b, Chris2015, Levinsen2015, Ardila2015, Volosniev2015, Grusdt2016, Grusdt2016Feb,
Shchadilova2016, Shchadilova2016b, Castelnovo2016, Ardila2016, Robinson2016, Jor2016, Hu2016, Rentrop2016, Lampo2017, Pastukhov2017, Yoshida2017, Guenther2018, Lingua2018},
as well as in solid state physics \cite{Landau48, Devreese2009, Alexandrov2009}, and mathematical physics \cite{Lieb1958, Lieb1997, Frank2010, Anapolitanos2013, Lim2018}.

We study the dynamics of the impurity within a BEC with an open quantum systems approach, namely we focus on the behavior of the former treating the latter as a mere source of noise and dissipation.
Very similar methods have been used recently to study 
a bright soliton in a superfluid in \cite{Efimikin2013}, a dark soliton in a BEC coupled to a non-interacting Fermi gas in \cite{Hurst2016}, the interaction between the components of a moving superfluid and the related collective modes \cite{Keser2016}, and an impurity in a Luttinger liquid in~\cite{Cugliandolo2012, Bonart2013}, or in a double-well potential \cite{Cirone2009, Haikka2011}. 
Particularly, in \cite{Lampo2017}, the dynamics of an impurity weakly interacting  with a homogeneous untrapped BEC \cite{Frolich1954,Alexandrov2009} were investigated by means of a paradigmatic model of open quantum system, the quantum Brownian motion (QBM) model. 
This model describes a particle that interacts with a thermal bath, made up by a huge number of harmonic oscillators, satisfying  Bosonic statistics \cite{GardinerBook, BreuerBook, SchlosshauerBook, Caldeira1983a, Caldeira1983b, Grabert1985,  Hu1992, Zurek2003, DeVega2017}.
In this framework, the impurity plays the role of the quantum Brownian particle and the bath is the set of excitations modes of the BEC.

In the present paper, we extend the "QBM point of view" developed in \cite{Lampo2017} to the situation where the BEC is trapped. 
We emphasize that it is of paramount importance to consider the scenario of a trapped BEC, as this way our model approaches the usual experimental set-up. 
In this case the gas results to be inhomogeneous in space, namely its density is space-dependent.
In particular, we consider a one-dimensional BEC trapped in a harmonic potential, yielding a parabolic density profile, \textit{i.e.} the Thomas-Fermi (TF) profile.  
Such a system has been already studied in \cite{Ohberg1997, Stringari1996, Petrov2004} in which the analytical form of the spectrum of the Bogoliubov excitations has been derived. 
We exploit this result to show that the Hamiltonian of the system may be written as that of the QBM model, where the impurity-bath interaction exhibits a non-linear dependence on the position of the former. 
Nevertheless, we find that for realistic experimental conditions indeed this reduces to the usual one of the QBM model, linearly dependent on the position.

From the QBM Hamiltonian, we derive the quantum Langevin equation describing the out-of-equilibrium dynamics of the impurity. The effect of the BEC then is manifested through the corresponding noise and damping terms present in this dynamical equation.
We solve the aforementioned equation and find the position and momentum variances for two distinct cases, (i) for an untrapped impurity and (ii) for a trapped impurity, where in this case we are referring to the impurity trap. In both of these cases, the gas remains confined in a harmonic potential.
In the untrapped case, the impurity does not reach equilibrium and shows a super-diffusive behavior at long-times. 
In the trapped case, the impurity reaches equilibrium in the long-time limit, and therefore the position and momentum variances reach stationary values. Interestingly, in this limit we find genuine position squeezing at low temperatures, which  can be enhanced with the coupling strength.   
The distinguishing difference with the homogeneous gas case, is that this coupling strength is now also a function of the gas trap frequency. As a result, we find that both the super-diffusion coefficient and the squeezing degree can be tuned with the BEC trap frequency, and we study this in detail. We emphasize that both squeezing and super-diffusion effects may be detected experimentally since they concern the position variance which constitutes a measurable quantity, as shown in \cite{Catani2012}.

Furthermore, the different form of the impurity-bath coupling constant leads to a new form of the spectral density (SD). This is a fundamental object in the open quantum systems framework, since it encodes all the relevant information concerning the effect of the environment on the impurity dynamics, once the degrees of freedom of the former are traced away. 
In particular we find that, although in both cases the SD shows a super-ohmic form ($J\sim\omega^\alpha,\quad\alpha>1$), the super-ohmic degree is higher when the medium is inhomogeneous, \textit{i.e. $\alpha_{\rm Inh}>\alpha_{\rm Hom}$}.
This suggests that the amount of memory effects carried out by the impurity dynamics is larger in the present situation.
A large part of the manuscript is devoted to evaluating in a quantitative manner the non-Markovian properties of the system. 
This kind of analysis is motivated by the recent efforts to understand the thermodynamical meaning of quantum non-Markovianity, and the attempts to employ such a feature as a resource to device new protocols for quantum technologies (see for instance~\cite{Bylicka2016}). 
In this context, the quantitative description of non-Markovianity for the polaron physics has never been examined properly: the only exception, at the best of our knowledge, is represented by \cite{Haikka2011}, although they consider an impurity embedded in a symmetric double well whose dynamics is treated by means of the spin-boson model, rather than the QBM one. 
Apart from the specific application to ultracold gases, it is important to note that the study of non-Markovianity for the QBM model has only been performed in \cite{Groblacher2015} where a measure based on the distance from the corresponding Lindblad map has been introduced, and in \cite{Vasile2011} which relies on a set of approximations that are not suitable to approach the polaron dynamics. 
We consider a number of techniques to investigate, in a formal manner, the non-Markovian character of the system. In all of these cases we find that for the inhomogeneous gas the non-Markovian degree is higher than in the homogeneous BEC.

The manuscript is organized as follows. In Sec.\ \ref{HamSec} we  derive the Hamiltonian of an impurity in a trapped BEC in the form  of the QBM model. In  Sec.\ \ref{HESec}, we write the quantum Langevin equation, derive the form of the SD, and find a general solution of the equation.  In Sec.\ \ref{PosVarSec} we solve this equation for the untrapped  (subsection \ref{NoImpTrapSec}) and trapped (subsection \ref{TrapSec}) impurity. 
In Sec.\ \ref{NMSec} we explore the non-Markovianity properties of the system employing (i) the measure introduced in \cite{Vasile2011}, (ii) the two-point correlation function, (iii) the distance with the ohmic process, and (iv) the evaluation of the back-flow of energy according the criterion presented in \cite{Guarnieri2016}. 
In Appendix\ \ref{ValLinApp} we discuss the validity of the linear approximation for the interacting Hamiltonian between the impurity and the BEC.   In Appendix \ref{LimitApp} we give a detailed discussion on the differences we found between the homogeneous and inhomogeneous BEC cases.

\section{Hamiltonian}\label{HamSec}
We consider an impurity with mass $m_{\rm I}$ embedded in a Bose-Einstein condensate with $N$ atoms  of mass $m_{\rm B}$. The system is described by the following Hamiltonian 
\begin{equation}\label{Ham}
H=H_{\rm I}+H_{\rm B}+H_{\rm BB}+H_{\rm IB},
\end{equation}
with
\begin{subequations}
\begin{align}
H_{\rm I}&=\frac{{\bk{p}}^2}{2m_{\rm I}}+U(\bk{r}),\label{H_I}\\
H_{\rm B}&=\int d^3 \bk{r}_{\rm B} \,\Psi^{\dag}(\bk{r}_{\rm B})\left[\frac{\bk{p}_{\rm B}^2}{2m_{\rm B}}+V(\bk{r}_{\rm B})\right]\Psi(\bk{r}_{\rm B}),\label{H_B}\\
H_{\rm BB}&=g_{\rm B}\int d^3\bk{r}_{\rm B}\, \Psi^\dag(\bk{r}_{\rm B})\Psi^\dag(\bk{r}_{\rm B})\Psi(\bk{r}_{\rm B}) \Psi(\bk{r}_{\rm B}),\label{H_BB}\\
H_{\rm IB}&= g_{\rm IB}\int d\bk{r}_{\rm B}\Psi^\dag(\bk{r}_{\rm B})\Psi(\bk{r}_{\rm B})\delta(\bk{r}-\bk{r}_{\rm B})\nonumber\\
&=g_{\rm IB}\Psi^\dag(\bk{r})\Psi(\bk{r}),\label{H_IB}
\end{align}
\end{subequations}
where $\bk{r}$ and $\bk{r}_{\rm B}$ denote the position operator of the impurity and the bosons, respectively. 
We assume contact interactions among the bosons and between the impurity and the bosons,  with strength given by the coupling constants $g_{\rm B}$ and $g_{\rm IB}$, respectively [see Eqs.\ \eqref{H_BB} and \ \eqref{H_IB}]. The impurity is trapped in a potential $U(\bk{r})=\frac{m_{\rm I}\Omega^2\bk{r}^2}{2}$. In this paper we discuss both the untrapped  ($\Omega=0$) and trapped cases  ($\Omega>0$).  The bosons are trapped in a harmonic potential, namely the potential in Eq.\ \eqref{H_BB} takes the form
\begin{equation}\label{HarmTrap}
V(\bk{r}_{\rm B})=\sum^3_{i=1}\frac{m_{\rm B}\omega^2_{\rm i} \left(r^{(i)}_{\rm B}\right)^2}{2}.
\end{equation}
This is the crucial difference with the analysis in Ref.~\cite{Lampo2017}, where the homogeneous BEC was discussed. The fact that the BEC is trapped gives rise to important consequences, both in the analytical derivation and in the results, as we will discuss throughout the rest of the paper. 

In this section we express the Hamiltonian\ \eqref{Ham} in the form of  the QBM model. 
We first write the field operator as the sum of the condensate state and the above-condensate part
\begin{equation}\label{BECdec}
\Psi=\Psi_0+\Psi',\quad \Psi_0\equiv\ave{\Psi}.
\end{equation}
We replace Eq.\ (\ref{BECdec}) in the Hamiltonian\ (\ref{Ham}) and make the BEC assumption, \textit{i.e.} that the condensate density greatly exceeds that of the above-condensate particles.
In particular this amounts to omitting the terms proportional to $\left(\Psi'\right)^3$, and $\left(\Psi'\right)^4$ in the resulting expressions. 
As shown in \cite{Ohberg1997}, one obtains
\begin{align}\label{MeanFieldGasHam}
&H_{\rm BB}+H_{\rm B}=H_{\rm 0}+\int d^3\bk{r}_{\rm B} \Psi'^\dagger H^{(\mathrm{sp})}_{\rm B}\Psi'\\
+&\frac{g_{\rm B}}{2}\left[4\md{\Psi_{\rm 0}}^2\Psi'^\dagger\Psi'+\Psi^2_{\rm 0}\Psi'^\dagger\Psi'^\dagger\right]
+\frac{g_{\rm B}}{2}\left(\Psi^*_{\rm 0}\right)^2\Psi'\Psi'\nonumber
,
\end{align}
with
\begin{equation}\label{H0Eq}
H_{\rm 0}\!=\!\!\int\!\! d^3\bk{r}_{\rm B}\Psi^\dagger_{\rm 0}(\bk{r}_{\rm B})\!\!\left[H^{(\mathrm{sp})}_{\rm B}+\frac{g_{\rm B}}{2}\md{\Psi_{\rm 0}(\bk{r}_{\rm B})}^2\right]\Psi_{\rm 0}(\bk{r}_{\rm B}),
\end{equation}
and
\begin{equation}
H^{(\mathrm{sp})}_{\rm B}\equiv\frac{\bk{p}^2_{\rm B}}{2m_{\rm B}}+V(\bk{r}_{\rm B}),
\end{equation}
is the single-particle gas Hamiltonian [see Eq.\ (\ref{H_B})].  
Note that in Eqs.\ \eqref{MeanFieldGasHam} and \eqref{H0Eq} we omitted the explicit dependence on $\bk{r}_{\rm B}$ to make the notation lighter.
Proceeding in a similar manner with the impurity-gas interaction, Eq.\ \eqref{H_IB}, one gets
\begin{align}
H_{\rm IB}=&g_{\rm IB}\left[\Psi^\dagger_{\rm 0}(\bk{r})+\Psi'^\dagger(\bk{r})\right]
\left[\Psi_{\rm 0}(\bk{r})+\Psi(\bk{r})\right]\nonumber\\
=&g_{\rm IB}\left[\md{\Psi_{\rm 0}(\bk{r})}^2+\Psi'^\dagger(\bk{r})\Psi_{\rm 0}(\bk{r})
+\Psi'(\bk{r})\Psi_{\rm 0}^\dagger(\bk{r})\right]\label{HintExp}
\end{align}  
where the term proportional to the square power of the above-condensate state has been neglected. 

In the QBM Hamiltonian, the environment is modeled as a set of {\it uncoupled} oscillators. To establish the analogy between the QBM Hamiltonian and that of the impurity immersed in a BEC, we diagonalize the part of the gas Hamiltonian, Eq.\ \eqref{MeanFieldGasHam}, to express it as a set of uncoupled  modes. 
With the Bogoliubov transformation
\begin{equation}\label{BogoliubovExp}
\Psi'(\bk{r}_{\rm B})=\sum_\nu\left[u_\nu(\bk{r}_{\rm B})b_\nu-v^*_\nu(\bk{r}_{\rm B})b^\dagger_\nu\right],
\end{equation}
one gets to the diagonalized Hamiltonian
\begin{equation}\label{BogModesHam}
H_{\rm B}+H_{\rm BB}=H_{\rm 0}+\sum_{\nu} E_{\nu} b^\dagger_{\nu} b_{\nu}, 
\end{equation}
where $E_\nu$ is the energy of the Bogoliubov excitations, which constitute the oscillating modes of the environment dressing the impurity, and $b^\dagger$ $(b)$ the related creation (annihilation) operators of these modes. Under the Bogoliubov transformations in Eq.\ \eqref{BogoliubovExp} the interaction Hamiltonian, Eq.\ \eqref{H_IB}, reads
\begin{align}\nonumber
H_{\rm IB}=&g_{\rm IB}\left[\sqrt{n_0(\bk{r})}\sum_{\nu} \left[u^*_{\nu}(\bk{r})-v^*_{\nu}(\bk{r})\right]b_{\nu}^\dagger+\text{c.c.}\right]\\
\equiv &g_{\rm IB}\left[\sqrt{n_0(\bk{r})}\sum_{\nu} f_{(\nu,-)} b_{\nu}^\dagger+\text{c.c.}\right]\label{intHam1}
\end{align}
where we put $\Psi_{\rm 0}\approx \sqrt{n_{\rm 0}}$.

To obtain the complete form of the Hamiltonian we need the expressions of the functions $u_\nu$ and $v_\nu$ introduced in Eq.~\eqref{BogoliubovExp}, as well as of the energy modes in Eq.\ \eqref{BogModesHam}.
An important difference with the homogeneous case is that, for the trapped BEC,  they have to be obtained as the eigenvectors and eigenvalues of the matrix associated to the Bogoliubov-de-Gennes (BdG) equations
\begin{subequations}
\label{BdG}
\begin{align}
&\label{BdG1}H^{(\mathrm{sp})}_{\rm B}u_\nu+g_{\rm B}n_{\rm 0}\left(2u_{\nu}-v_{\nu}\right)=\left(\mu+E_{\nu}\right)u_{\nu}\\
&\label{BdG2}H^{(\mathrm{sp})}_{\rm B}u_{\nu}+g_{\rm B}n_{\rm 0}\left(2u_{\nu}-v_{\nu}\right)=\left(\mu-E_{\nu}\right)u_{\nu}.
\end{align}
\end{subequations}
The solutions of the BdG equations satisfy the orthogonality condition
\begin{equation}
\int d\bk{r}\left(u_\nu u^*_{\nu'}-v_\nu v^*_\nu\right)=\delta_{\nu\nu'}.
\end{equation}
In general, the solution of the BdG equations~(\ref{BdG}) does not constitute a simple problem, and often requires the employment of numerical methods. 
For a BEC confined in one dimension and  in the TF limit, one can solve them analytically as shown in~\cite{Petrov2004}. 
In the current work we focus exactly on the aforementioned situation, namely a gas confined in one dimension  with a TF density profile
\begin{equation}\label{TFDensProf}
n_0(x)=\frac{\mu}{g_{\rm B}}\left(1-\frac{x^2}{R^2}\right),\quad R=\sqrt{2\mu/m_{\rm B}\omega^2_{\rm B}},
\end{equation}
where  $\omega_{\rm B}$ is the trapping frequency in the direction $x$ [see Eq.\ \eqref{HarmTrap}]. Here,  $R$ is the TF radius and the chemical potential is
\begin{equation}
\mu=\left(\frac{3}{4\sqrt{2}}g_{\rm B}N\omega_{\rm B}\sqrt{m_{\rm B}}\right)^{2/3}. 
\end{equation}   
Then, the solution of  the BdG equations~(\ref{BdG}) gives the following spectrum
\begin{equation}\label{BogSpec}
E_{j}=\hbar\omega_{\rm B}\sqrt{j(j+1)}\equiv\hbar\omega_{j}, 
\end{equation}
with corresponding Bogoliubov modes 
\begin{align}\label{fMin1D}
f_{\rm (j,-)}=\sqrt{\frac{j+1/2}{R}}\sqrt{\frac{2\mu}{E_{j}}\left[1-\left(\frac{x}{R}\right)^2\right]}L_{j}\left(x/R\right).
\end{align}
where $L_{j}(z)$ represent the Legendre polynomials and $j$ is the integer quantum number labeling the spectrum. 

Finally, we replace the expressions of the Bogoliubov modes, Eq.\ \eqref{fMin1D}  in Eq.\ \eqref{intHam1} to get  the Hamiltonian of an impurity embedded in a BEC in 1D with a TF density profile, 
\begin{equation}
\label{analogousQBM}
H=H_{\rm I}+H_{\rm E}+H_{\rm int},
\end{equation}
with
\begin{equation}
H_{\rm E}=\sum_{j}E_{j}b^{\dagger}_{j}b_{j}, 
\end{equation}
and
\begin{align}\label{generalIntHam}
H_{\rm int}&=\sum_{j}g_{\rm IB}\sqrt{n_{\rm 0}(x)}f_{\rm (j,-)}(x)\left(b_{j}+b^\dagger_{j}\right)\nonumber\\
&\equiv\sum_{j}F_{j}(x)\left(b_{j}+b^\dagger_{j}\right),
\end{align}
The  Hamiltonian\ \eqref{analogousQBM} is analogous to that of the QBM model, where one identifies the system Hamiltonian as $H_{\rm I}$, the environment set of oscillators as $ H_{\rm E}$, and the interaction between system and environment as $H_{\rm int}$. Notably, in our case, the latter presents a non-linear dependence on the position impurity.  
There is a number of existing techniques aimed at dealing with the QBM model with this kind of non-linearity.  
For instance, one could recall the master equation treatment in the Born-Markov regime in~\cite{Massignan2015}, or in the Lindblad framework~\cite{Lampo2016}. 
Beyond these approximations, one could also deal with this problem considering the non-linear Heisenberg equation obtained by such a non-linear interacting Hamiltonian, as in~\cite{Barik2005}. In this case one deals with a generalized Langevin equation with a state-dependent damping and a multiplicative noise. 
Moreover, there is the procedure presented in~\cite{Lim2018} relying on quantum stochastic calculation, valid for the small impurity mass limit. 

The problem in applying all these methods in our case, lies on the fact that the interaction Hamiltonian\ \eqref{generalIntHam} presents a dependence on the position that is different for a different $j$ index, \textit{i.e.} the impurity-bath coupling has a different form as a function of the impurity's position for bosons of different eigenmodes. 
To overcome this difficulty, we restrict ourselves to the regime constrained by the condition $x/R\ll1$, that is we study the dynamics of the impurity in the middle of the trap.   
Here, it is possible to expand the interaction term in Eq.\ \eqref{generalIntHam} at the first order in $x/R$
\begin{equation}\label{IntHamMiddle}
H_{\rm I}=\sum_{j}\hbar g_{j}x\left(b_{j}+b^\dagger_{j}\right),
\end{equation}
in which
\begin{align}\label{CoupConst}
g_{j}\!=\!\frac{g_{\rm IB}\mu}{\hbar\pi^{3/2}}\!\left[\!\frac{1+2j}{\hbar\omega_{\rm B}g_{\rm B}R^3}\!\right]^\frac{1}{2}\!\frac{\Gamma\left[\frac{1}{2}\left(1\!-\!j\right)\right]\Gamma\left[\frac{1}{2}\left(1\!+\!j\right)\right]\sin\left(\pi j\right)}{\left[j(j+1)\right]^{1/4}}.
\end{align}
This linear approximation above, is discussed in Appendix \ref{ValLinApp}. There we show that assuming that we are in the linear approximation regime  is appropriate for realistic values of the system parameters. 
The interaction Hamiltonian above shows a linear dependence on the positions of both the impurity and the oscillators of the bath. 
This is exactly the situation of the QBM model. 
Note that, contrary to the homogeneous gas, the coupling in this case is not to the momentum degree of freedom of the bath's harmonic oscillators but rather to their positions. This however, does not imply a qualitative change with respect to the homogeneous case, because the bath variables only play a role in the environmental self-correlation functions, which remain the same as those presented in~\cite{Lampo2017}. 

The substantial change with respect to the homogeneous medium is the new structure of the bath-impurity coupling constant in Eq.\ \eqref{CoupConst}.
Such a quantity exhibits a different dependence on the system parameters in comparison to that derived in the homogeneous case (see Eq.\ (42) of \cite{Lampo2017}).
In particular, we obtain now a dependence on the frequency of the gas trap, that may be tuned in order to modify the properties of the impurity. 
In the rest of the manuscript we shall discuss the effects of the new form of the bath-impurity coupling constant. 
We will see for instance that the different dependence on the bath index $j$ alters the amount of memory effects defining the non-Markovian properties of the system.

\section{Quantum Langevin Equation}\label{HESec}
After expressing  the Hamiltonian of the system in the form of the QBM one, we are now in the position to provide a careful quantitative description of the motion of the impurity using an open quantum systems approach. First, we write the Heisenberg equations 
\begin{align}
&\dot{x}(t)=\frac{i}{\hbar}\comm{ {H}}{ {x}(t)},\quad\dot{p}(t)=\frac{i}{\hbar}\comm{ {H}}{ {p}(t)},\label{EqXP}\\
&\dot{b}_k(t)
=\frac{i}{\hbar}\comm{ {H}}{ {b}_k(t)},\quad\dot{ {b}}^\dagger_k(t)=\frac{i}{\hbar}\comm{ {H}}{ {b}^\dagger_k(t)}.\label{Eqbdag}
\end{align}
These equations may be combined according the procedure presented in \cite{BreuerBook, Lampo2017} to derive an equation for the position impurity in the Heisenberg picture,
 \begin{equation}\label{EqDiffFin}
\ddot{x}(t)+\Omega^2 x(t)+\der{}{t}\int^{t}_0\Gamma(t-s) x(s)ds=\frac{B(t)}{m_{\rm I}}.
\end{equation}
Such an equation is formally identical to the Langevin one derived in the context of classical Brownian motion, and
completely rules the temporal evolution of the impurity motion. 
At this level, the influence of the environment is contained in the term in the right hand-side
\begin{equation}
B(t)=\sum_{j} \hbar g_{ j}( {b}^{\dagger}_{ j}e^{-i\omega_{ j}t}+ {b}_{ j}e^{+i\omega_{ j}t}),
\end{equation}
which plays the role of the stochastic noise, and in the damping kernel
\begin{equation}
\label{DampingKernel}
\Gamma(\tau)=\frac{1}{m_{\rm I}}\int^{\infty}_0\frac{J(\omega)}{\omega}\cos(\omega\tau)d\omega,
\end{equation}
where we introduced the spectral density (SD), defined as
\begin{equation}\label{SDdef}
J(\omega)=\sum_{k\neq0}\hbar g^2_{k}\delta\pton{\omega-\omega_{k}}.
\end{equation}
The SD completely determines the form of the damping kernel. 
This is also true for the noise term, since it fulfills the relation
\begin{equation}\label{NoiseCorr}
\ave{\{B(s),B(\sigma)\}}=2\hbar\nu(s-\sigma),
\end{equation}
in which
\begin{equation}\label{NoiseDamp}
\nu(\tau)=\int^{\infty}_{0}J(\omega)\coth\left(\frac{\hbar\omega}{2k_{\rm B}T}\right)\cos\left(\omega\tau\right)d\omega
\end{equation}
is the noise kernel. 

Therefore, the influence of the environment on the impurity motion is completely determined once the form of the SD is determined. 
From Eq.\ \eqref{SDdef}, we see that the SD is determined by the coupling constant whose form is given in Eq.\ \eqref{CoupConst}.
Replacing this quantity in Eq.\ \eqref{SDdef} and turning
the discrete sum in $j$ into a continuous variable integral, one gets
\begin{align}\nonumber
J(\omega)&=\frac{2g_{\rm IB}^2\mu^2}{g_{\rm B}R^3(\hbar\omega_{\rm B})^2}\left(\frac{\omega}{\omega_{\rm B}}\right)^4\theta\left(\omega-\omega_{\rm B}\right)\\\label{spectralDensity}
&\equiv m_{\rm I}\gamma\frac{\omega^4}{\Lambda^3}\theta\left(\omega-\Lambda\right),
\end{align}
with
\begin{equation}
\gamma=\frac{2g_{\rm B}}{m_{\rm I}\omega_{\rm B}R^3}\left(\frac{\eta\mu}{\hbar\omega_{\rm B}}\right)^2,\quad\eta=\frac{g_{\rm IB}}{g_{\rm B}},\quad\Lambda=\omega_{\rm B}.
\end{equation}
In Eq.\ \eqref{spectralDensity}, $\theta\left(\omega-\Lambda\right)$ is the Heaviside step function, representing an ultraviolet cut-off, that has been put {\it ad-hoc} in order to regularize the divergent character of the SD at high-frequency. 
This, however, does not play any role in the dynamics of the system at long-times, as nor the presence, neither the form, of the cut-off affects the dynamics of the impurity at long times. This can be shown by recalling the Tauberian theorem \cite{FloydBook, FellerBook}. 

Therefore, in the middle of the trap ($x\ll R$) and at long times ($\omega\ll\omega_{\rm B}$) we obtain a super-Ohmic SD. 
This form of the SD implies the presence of memory effects in the dynamics of the system. 
In fact, only if the damping kernel reduces to a Dirac Delta, Eq.\ \eqref{EqDiffFin} acquires a local-in-time structure, making the evolution of the impurity's position independent of its past history. 
Indeed, by replacing the SD\ \eqref{spectralDensity} in the definition of the damping kernel, Eq.\ \eqref{DampingKernel},  one gets
\begin{align}\label{dampExp}
\Gamma(t)=&\frac{\gamma\left[6+3\left(\omega^2_{\rm B}t^2-2\right)\cos\left(\omega_{\rm B} t\right)\right]}{t^4\omega^3_{\rm B}}\\
+&\frac{\gamma \omega_{\rm B} t\left[\left(\omega_{\rm B}t\right)^2-6
\right]\sin\left(\omega_{\rm B} t\right)}{t^4\omega^3_{\rm B}}.\nonumber
\end{align}
The form of the damping kernel presented above shows that Eq.\ \eqref{EqDiffFin} is non-local-in-time and the dynamics of the impurity carries a certain amount of memory effects.
We underline here an important difference with the case in which the BEC is untrapped: in that situation the SD is proportional to the third power of the frequency~\cite{Lampo2017}, while now it goes as the fourth one. 
We conclude that the presence of the trap for the gas increases the super-Ohmic degree and changes the details of the derivation to be developed below, in comparison with the homogeneous case. 
In Appendix \ref{LimitApp} we show that the difference in one power of $\omega$ between the SD for an homogeneous and inhomogeneous BEC parallels the different behavior of the density of states in both cases.     Apart from the technical details of the calculations, the higher super-Ohmic degree alters the amount of memory effects characterizing the system dynamics. 
The difference between this aspect in the homogeneous and inhomogeneous case will be treated in the last part of the work.
This is a consequence of the different structure of the coupling constant presented in Eq.~\eqref{CoupConst} and, in particular, of its dependence on the bath index $j$. 
The new form of the coupling constant does not affect only the analytical profile of the SD in the frequency domain, but also its prefactor $\gamma$,  termed \textit{damping constant}, which is related to the timescale of the dissipation process. This new form of the damping constant depends on the frequency of the trap of the gas, and interestingly this may be tuned in order to modify the qualitative properties of the solution of Eq.~\eqref{EqDiffFin}, as we will show in the next part of the manuscript.

The solution of Eq.\ \eqref{EqDiffFin} is
\begin{equation}\label{XHeis}
x(t)\!=\!G_{\rm 1}(t)x(0)+G_{\rm 2}(t)\dot{x}(0)+\frac{1}{m_{\rm I}}\int^{t}_0G_{\rm 2}(t-s)B(s)ds,
\end{equation} 
where the functions $G_{\rm 1}$ and $G_{\rm 2}$ are defined through their Laplace transforms 
\begin{align}
&\mathcal{L}_z[G_{\rm 1}(t)]=\frac{z+\mathcal{L}_z[\Gamma(t)]}{z^2+\Omega^2+z\mathcal{L}_z[\Gamma(t)]},\label{LTG1}\\
&\mathcal{L}_z[G_{\rm 2}(t)]=\frac{1}{z^2+\Omega^2+z\mathcal{L}_z[\Gamma(t)]},\label{LTG2}
\end{align}
and satisfy
\begin{align}
&G_{\rm 1}(0)=1,\quad \dot{G}_{\rm 1}(0)=0\label{InCondG1},\\
&G_{\rm 2}(0)=0,\quad \dot{G}_{\rm 2}(0)=1\label{InCondG2}.
\end{align}
The Laplace transform of the damping kernel is what carries out the properties of the environment in the solution of the position impurity equation. 
Recalling the definition of the damping kernel we find
\begin{align}
\mathcal{L}_z[\Gamma(t)]&=\frac{1}{m_{\rm I}}{\int_0^{\infty}} dt e^{-zt}\cos(\omega t){\int_0^{\infty}} d\omega J(\omega)/\omega\nonumber\\
&=\frac{z\gamma}{\omega^3_{\rm B}}{\int_0^{\omega_{\rm B}}}d\omega\frac{\omega^3}{\omega^2+z^2}\label{LTDampCalc},
\end{align}
where we used the expression of the SD in Eq.\ \eqref{spectralDensity} and the formula for the Laplace transform of the cosine
\begin{equation}
\int^\infty_0  e^{-zt}\cos(\omega t)dt=\frac{z}{\omega^2+z^2}.
\end{equation}
The integral\ \eqref{LTDampCalc} may be calculated straightforwardly noting that
\begin{align}
&{\int_0^{\omega_{\rm B}}}\frac{\omega^3}{\omega^2+z^2}d\omega={\int_0^{\omega_{\rm B}}}\omega\left(1-\frac{z^2}{\omega^2+z^2}\right)d\omega\nonumber\\
=&\frac{1}{2}\left[\omega^2_{\rm B}+z^2\log\left(\frac{z^2}{z^2+\omega^2}\right)\right]\label{LTDampCalc1}.
\end{align}
In the end, replacing Eq.\ \eqref{LTDampCalc1} into Eq.\ \eqref{LTDampCalc}, we obtain
\begin{equation}\label{LTdamp}
\mathcal{L}_z[\Gamma(t)]=\frac{z\gamma}{2\omega^3_{\rm B}}\left(\omega^2_{\rm B}+z^2\log\left[\frac{z^2}{z^2+\omega^2_{\rm B}}\right]\right).
\end{equation}  
Such a quantity completely fixes the kernels in Eqs.\ \eqref{LTG1} and \eqref{LTG2} and thus the temporal evolution of the impurity position in the Heisenberg picture. 
The problem of deriving an explicit expression for it reduces now to the inversion of the Laplace transform in Eqs.\ \eqref{LTG1} and \eqref{LTG2}.   

\section{Position variance}\label{PosVarSec}
The motion of the impurity is described by the second-order stochastic equation of the Langevin type\ \eqref{EqDiffFin}.
We proceed now to solve this equation in order to evaluate the position variance, which constitutes a measurable quantity~\cite{Catani2012}. 
For this goal we distinguish two situations: the case where there is no  trap for the impurity [$\Omega=0$ in Eq.\ \eqref{H_I}], and that in which there is a harmonic trap ($\Omega>0$). 
We remark once more that in both situations the gas is harmonically trapped, \textit{i.e.} $\omega_{\rm B}>0$. 

\begin{figure}
\begin{center}
\includegraphics[width=0.95\columnwidth]{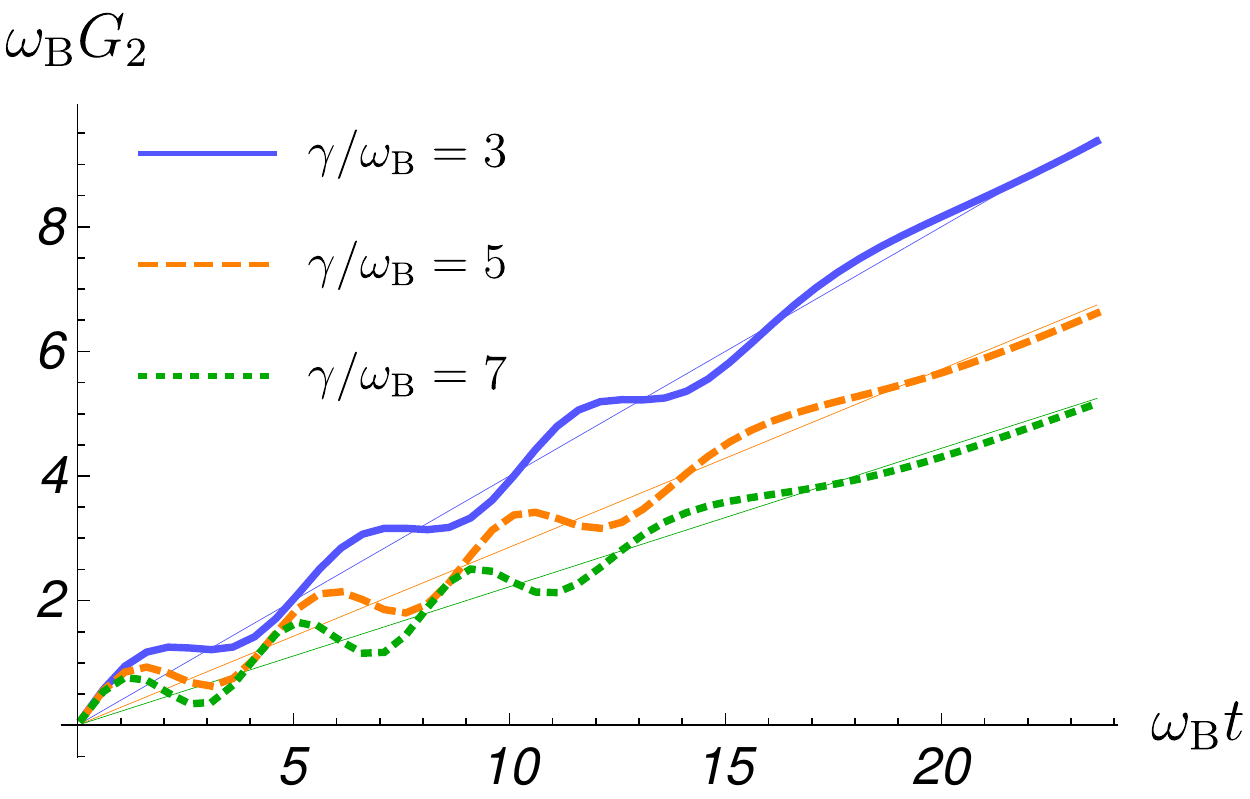}
\caption{\label{G2num} Time-dependence of the function $G_{2}$, defined through its Laplace transform in Eq.\ \eqref{LTG2}. 
The thick lines represent the numerical calculation with the Zakian algorithm, while the corresponding continue thin ones
 refer to the expression in Eq.\ \eqref{G2An}, valid in the long-time limit. 
}
\end{center}
\end{figure}

\subsection{Untrapped impurity}\label{NoImpTrapSec}
In Sec.\ \ref{HESec} we showed that the problem of solving Eq.\ \eqref{EqDiffFin} reduces to that of inverting the Laplace transforms\ \eqref{LTG1} and \eqref{LTG2}. 
The former may be inverted immediately since, when $\Omega=0$, it takes the form
\begin{equation}
\mathcal{L}_z[G_{\rm 1}(t)]=1/z,
\end{equation}
and so
\begin{equation}
G_{\rm 1}(t)=1.\label{G1Un}
\end{equation}
This result holds regardless of the properties of the environment, namely for any SD, and in fact corresponds to that derived in the homogeneous gas. 

The situation is different for Eq.\ \eqref{LTG2}, where the properties of the environment play a crucial role since they enter through the damping kernel.
Here, one cannot perform the inversion of the Laplace transform analytically due the presence of the logarithm [see Eq.\ \eqref{LTdamp}].  
Therefore, we recall the Zakian numerical method, discussed in~\cite{Wang2015}. 
Such a method relies on the fact that the inverse Laplace transform $f(t)$ of a function $F(z)$ is approximated as
\begin{equation}\label{Zakian}
\tilde{f}(t)=\frac{2}{t}\sum^{N}_{j=1}\text{Re}\left[k_{\rm j}F\left(\alpha_{\rm j}/t\right)\right],
\end{equation} 
with $\alpha_{\rm j}$ and $k_{\rm j}$  constants that can be either complex or reals. 

The expression of $G_{\rm 2}$ as a function of time is presented in Fig.\ \ref{G2num}. 
The kernel shows an oscillating behavior that diverges linearly in the long-time regime. 
Such a long-time limit corresponds to $\text{Re}[z]\ll\omega_{\rm B}$, where the logarithm in the Laplace transform of the damping kernel, \textit{i.e.} the second term in the right hand-side of Eq.\ \eqref{LTdamp}, is negligible. 
If we keep only the linear term in $z$ within such an equation it is possible to find an explicit analytical expression for the Laplace transform of $G_{\rm 2}$,
\begin{equation}
\mathcal{L}_z[G_{\rm 2}(t)]=\frac{1}{z^2(1+\frac{\gamma}{2\omega_{\rm B}})},
\end{equation}
that can be easily inverted
\begin{equation}\label{G2An}
G_{\rm 2}=\frac{t}{1+\frac{\gamma}{2\omega_{\rm B}}}\equiv\frac{t}{\tilde{\alpha}}. 
\end{equation}
This expression represents the long-time behavior of $G_{\rm 2}$ and is plotted in Fig.\ \ref{G2num} for different values of the damping (dashed lines). 
The figure shows the agreement between the numerical solution and the long-time analytical one.

\begin{figure}
\begin{center}
\includegraphics[width=0.95\columnwidth]{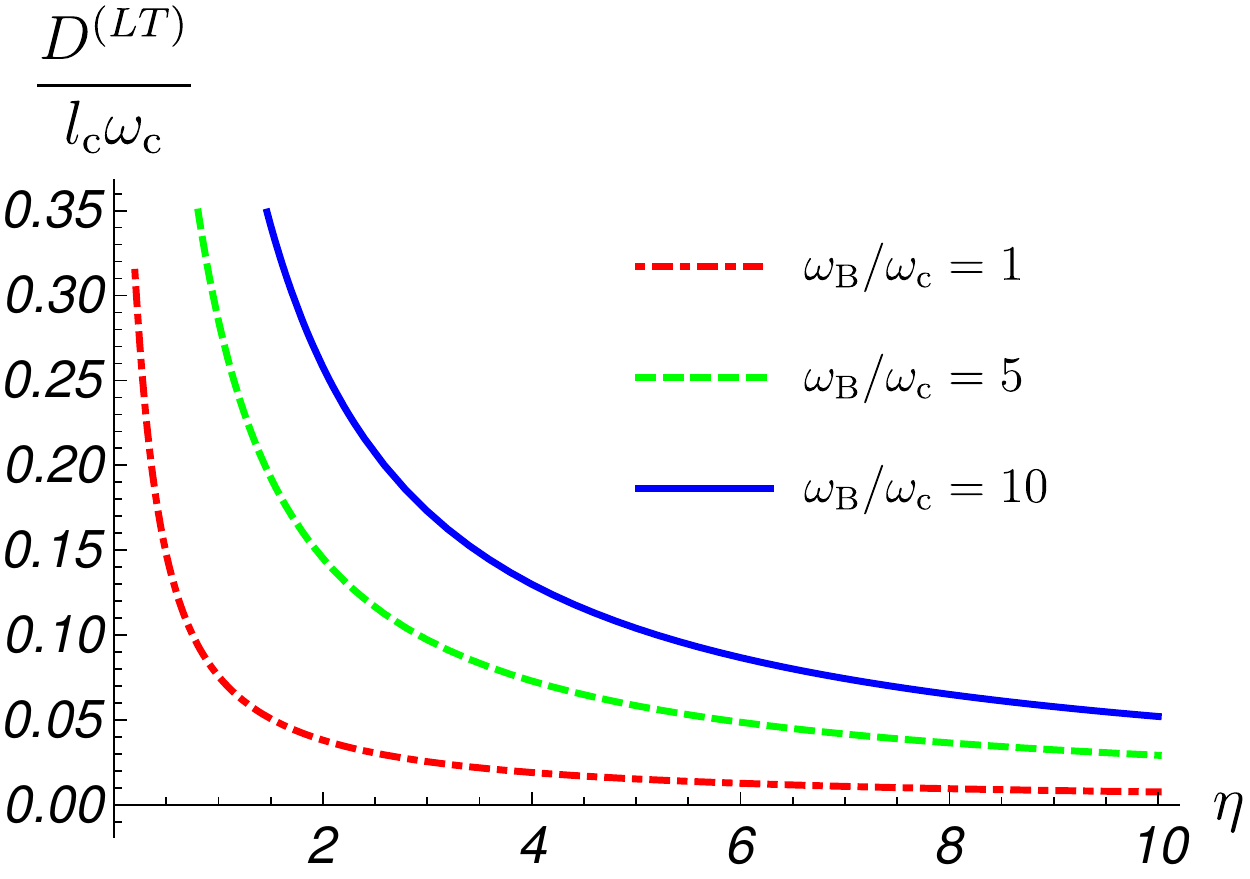}
\caption{\label{diffPlot} Super-diffusion coefficient in Eq.\ \eqref{SuperDiffLT} as a function of the interaction strength for different values of the gas trap frequency. 
We present the results for an impurity of Yb embedded in a Rb gas of $N=50000$ atoms with   coupling strength $g_{\rm B}=10^{-38}\text{J}\cdot\text{m}$. 
In this context the units of frequency are $\omega_{\rm c}=\frac{m_{\rm I}g^2_{\rm B}}{\hbar^3}$, while the units of the length are $l_{\rm c}=\frac{\hbar^2}{m_{\rm I} g_{\rm B}}$.  }
\end{center}
\end{figure}

The knowledge of $G_{\rm 1}$ and $G_{\rm 2}$ fixes the structure of the impurity position operator, providing a description of the motion of the particle. 
The expression for $G_{\rm 2}$ in Eq.\ \eqref{G2An} induces a ballistic term in the time-evolution of the impurity position. 
This means that the impurity runs-away from its initial position. 
Such a behavior can be characterized in a quantitative manner by means of the position variance.
Actually, rather than the position variance we employ a physically equivalent object called mean-square-displacement (MSD), defined as
\begin{equation}
\mbox{MSD}(t)=\ave{\left[x(t)-x(0)\right]^2},
\end{equation}
which provides the deviation between the position at time $t$ and the initial one.  
In the long-time limit it is possible to write
\begin{align}\label{d(t)1}
&\mbox{MSD}(t)=\left(\frac{t}{\tilde{\alpha}}\right)^2\ave{ \dot{x}(0)^2}\\ \nonumber
+&\frac{1}{2\left(\tilde{\alpha} m_{\rm I}\right)^2}\!\int^t_0\!ds\int^t_0\!d\sigma(t-s)(t-\sigma)\ave{\{B(s),B(\sigma)\}},
\end{align}
where we considered a factorizing initial state $\rho(t)=\rho_{\rm S}(0)\otimes\rho_{\rm B}$. The initial conditions of the impurity and  bath oscillators are then uncorrelated. Then,  averages of the form $\ave{\dot{x}(0)B(s)}$ vanish.  The integral in the second line of Eq.\ \eqref{d(t)1} can be solved recalling the expression for the two-time correlation function of the noise term\ \eqref{NoiseCorr} and that for the noise kernel\ \eqref{NoiseDamp}.  
Here, the hyperbolic cotangent can be approximated in two limits: (i) in the zero-temperature limit, where it can be approximated to one; and (ii) in the high-temperature limit, where it can be approximated to the inverse of its argument. In these two limits we have, respectively,
\begin{align}
&\text{MSD}^{\rm{(LT)}}(t)=\left[\ave{ \dot{x}(0)^2}+\frac{\hbar\gamma}{3m_{\rm I}}\right]\left(t/\tilde{\alpha}\right)^2,\label{MSDLT}\\
&\text{MSD}^{\rm{(HT)}}(t)=\left[\ave{ \dot{x}(0)^2}+\frac{k_{\rm B}T\gamma}{m_{\rm I}\omega_{\rm B}}\right]\left(t/\tilde{\alpha}\right)^2\label{MSDHT}. 
\end{align}
In both cases, the MSD is proportional to the square of time. 
This is a consequence of the super-Ohmic form of the SD, and can be considered as a witness of memory effects. 
The dependence on time is the same as for the homogeneous case. This is due to the fact that, in the long-time limit, the damping kernel and hence $G_{\rm 2}$ approaches the same function. 
Most importantly, for a trapped BEC the diffusion coefficients exhibit a different dependence on the system parameters. This is very relevant for the experimental validation of the current theory. 
In Fig.\ \ref{diffPlot} we plot the super-diffusion coefficient
\begin{equation}\label{SuperDiffLT}
D^{\rm{(LT)}}=\frac{\hbar\gamma}{3m_{\rm I}\tilde{\alpha}}
\end{equation}
related to the MSD in the low-temperature limit. 
Such a coefficient can be interpreted as the average of the square of the speed with which the impurity runs away. 
The picture shows that the quantity in Eq.\ \eqref{SuperDiffLT} decreases as the interaction strength grows. This implies that the gas acts as a damper on the motion of the impurity.  
Surprisingly, the value of the super-diffusion coefficients takes larger values as the gas trap frequency grows. 
One has to note that, as $\omega_{\rm B}$ grows, the density of the gas increases as well, and therefore the number of collisions yielding the Brownian motion also grows. 
The study of the super-diffusion coefficient at high-temperature shows the same behavior.

\subsection{Harmonically trapped impurity}\label{TrapSec}
We  now study the dynamics of the impurity when it is externally trapped, \textit{i.e.} we look into the case in which $\Omega>0$. 
In this case the inversion of the Laplace transforms constitutes a difficult task and it is not immediate to get an analytical explicit expression even at long-time. 
We proceed by employing the numerical Zakian method introduced above. 
\begin{figure}
\begin{center}
\includegraphics[width=0.95\columnwidth]{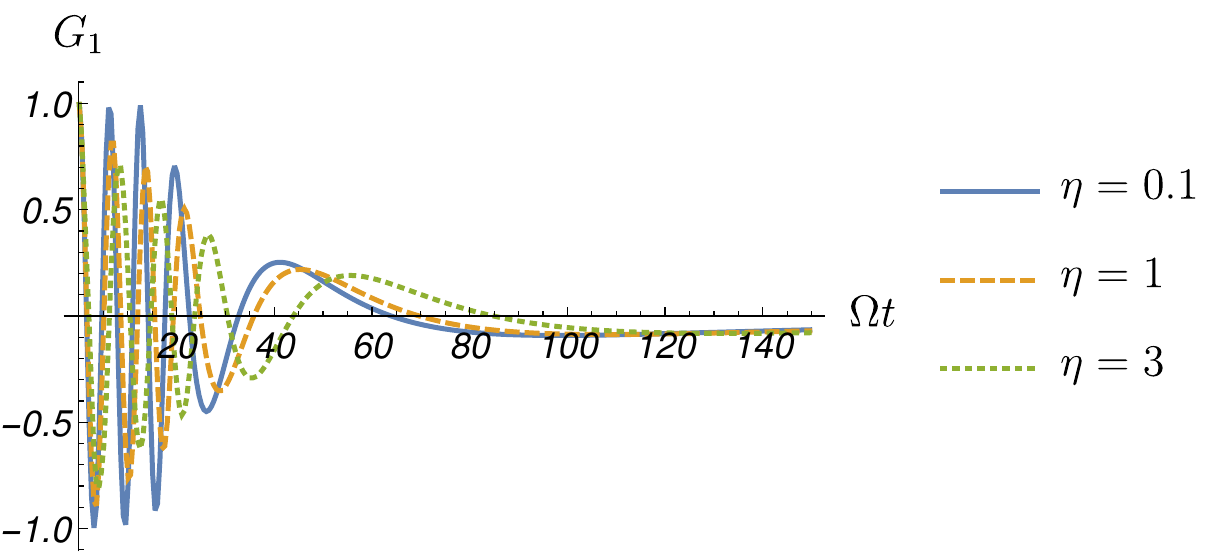}
\includegraphics[width=0.95\columnwidth]{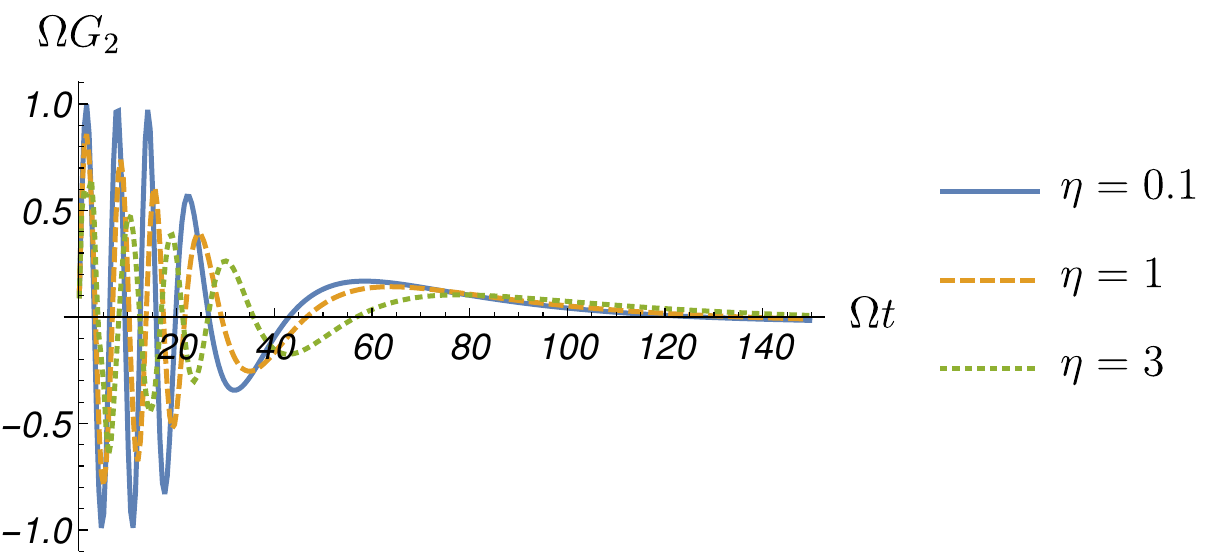}
\caption{\label{G1-2Trap} Time-dependence of the function $G_{1}$ (top) and  $G_{2}$ (bottom), defined through the Laplace transforms in Eqs.\ \eqref{LTG1} and \eqref{LTG2}, respectively. 
The plots refer to an impurity of Yb in a trap with a frequency $\Omega=2\pi\cdot200$ Hz, embedded in a Rb gas of $N=5000$ atoms with  trap frequency $\omega_{\rm B}=2\pi\cdot800$ Hz and    coupling strength $g_{\rm B}=0.5\cdot10^{-37}$ J$\cdot$m. }
\end{center}
\end{figure}
In Fig.\ \ref{G1-2Trap} we show the functions $G_{\rm 1}$ and $G_{\rm 2}$, where one can observe an oscillating behavior in both cases, which gets damped for long times. 
This damping of the oscillation implies that the contribution of the initial condition vanishes in the long-time limit. Also, this damping implies that the impurity reaches an equilibrium state where it sits on average on the center of the trap, and  its position and momentum variances  are independent of time. 
Thus, in the long-time limit, the variances can be represented by
\begin{align}
&\ave{x^2}=\frac{\hbar}{2\pi}\int^{+\omega_{\rm B}}_{-\omega_{\rm B}}d\omega\coth\left(\hbar\omega/2k_{\rm B}T\right)\tilde{\chi}''(\omega),\label{X2Trap2}\\
&\ave{p^2}=\frac{\hbar m^2_{\rm I}}{2\pi}\int^{+\omega_{\rm B}}_{-\omega_{\rm B}}\omega^2d\omega\coth\left(\hbar\omega/2k_{\rm B}T\right)\tilde{\chi}''(\omega),\label{P2Trap2}
\end{align}
where
\begin{equation}\label{ResponseFunction}
\tilde{\chi}''(\omega)=\frac{1}{m_{\rm I}}\frac{\zeta(\omega)\omega}{\left[\omega\zeta(\omega)\right]^2+
\left[\Omega^2-\omega^2+\omega\theta(\omega)\right]^2},
\end{equation}
is the response function, and
\begin{align}\label{RealAndImLTDamp}
\zeta\left(\omega\right)=\text{Re}\{\mathcal{L}_{\tilde{z}}\left[\Gamma(t)\right]\},\quad\theta\left(\omega\right)=\text{Im}\{\mathcal{L}_{\tilde{z}}\left[\Gamma(t)\right]\}. 
\end{align}
with $\tilde{z}=-i\omega+0^{+}$. 
The expression in Eq.\ \eqref{X2Trap2} can be obtained directly by the solution of the Heisenberg equations in Eq.\ \eqref{XHeis}, according the procedure presented in \cite{Lampo2017}, and corresponds to the contribution provided by the stochastic noise. 

We next study the dependence of the position and momentum variances, Eqs.\ \eqref{X2Trap2} and \eqref{P2Trap2},  on the system parameters, such as temperature and coupling strength. These parameteres can be tuned in experiments. 
To this end, we recall the dimensionless variables
\begin{equation}\label{DLessVar}
\delta_x=\sqrt{\frac{2m_{\rm I}\Omega\ave{x^2}}{\hbar}},\quad\delta_p=\sqrt{\frac{2\ave{p^2}}{m_{\rm I}\hbar\Omega}},
\end{equation}
in terms of which the Heisenberg principle reads as $\delta_x\delta_p\geq1$. 
Note that the evaluation of the variances in Eq.\ \eqref{DLessVar} relies on the calculation of the integrals\ \eqref{X2Trap2} and \eqref{P2Trap2}.
Similar integrals also appear in \cite{Lampo2017}, where they have been solved analytically by recalling the Residuous theorem. 
For this goal, one needs to cast the denominator in Eq.\ \eqref{ResponseFunction} in a polynomial form and so expand the Laplace transform of the damping kernel in Taylor powers. 
It is possible to show that in the inhomogeneous case, even by performing such an expansion in $z/\omega_{\rm B}$ a logarithm survives, and the denominator in Eq.~\eqref{ResponseFunction} cannot be reduced to a polynomial. 
Accordingly the integrals~\eqref{X2Trap2} and~\eqref{P2Trap2} cannot be solved analytically and one has to proceed numerically. 
Note also that such a numerical evaluation deserves to be performed carefully since the response function\ \eqref{ResponseFunction} is strongly narrowed around $\omega\approx\Omega$ and this affects the convergence of the integral. 
One has therefore to properly tune the number of recursive subdivisions and the number of effective digits of precision should be sought in the final result.

\begin{figure}
\begin{center}
\includegraphics[width=0.95\columnwidth]{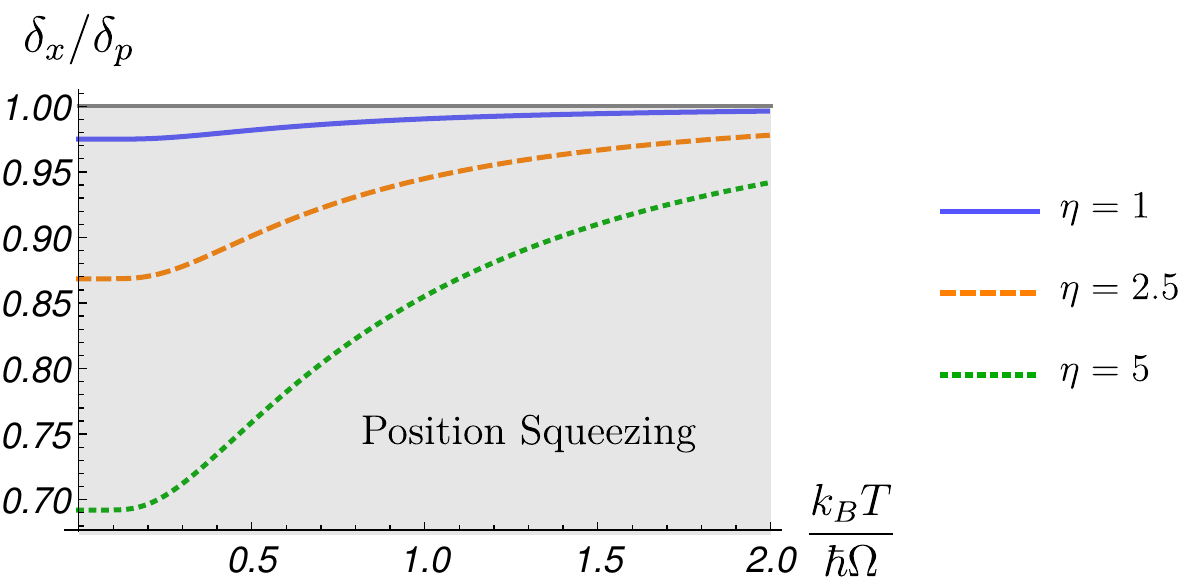}
\caption{\label{uncertEllipse} 
Temperature dependence of the ratio $\delta_p/\delta_p$ between the variances introduced in Eq.\ \eqref{DLessVar}. 
The plot refers to an impurity of Yb in a trap with a frequency $\Omega=2\pi\cdot50$ Hz, embedded in a Rb gas of $N=5000$ atoms  with  trap frequency $\omega_{\rm B}=2\pi\cdot500$ Hz and    coupling strength $g_{\rm B}=0.6\cdot10^{-38}$ J$\cdot$m. }
\end{center}
\end{figure}

In Fig.\ \ref{uncertEllipse} we study the behavior of the ratio $\delta_x/\delta_p$ as a function of the temperature for different values of the coupling strength. 
This gives the eccentricity of the uncertainty ellipse.
Such an ellipse takes the form of a circle at high-temperature, \textit{i.e.} $\delta_x\approx\delta_p$, for different values of the coupling strength. 
Precisely, it approaches the circular Gibbs-Boltzmann distribution with $\delta_x=\delta_p\sim\sqrt{T}$. 
At low temperature, instead, the uncertainties ellipse exhibits position squeezing ($\delta_x<\delta_p$), that is enhanced as the coupling strength increases. 
\begin{figure}
\begin{center}
\includegraphics[width=0.95\columnwidth]{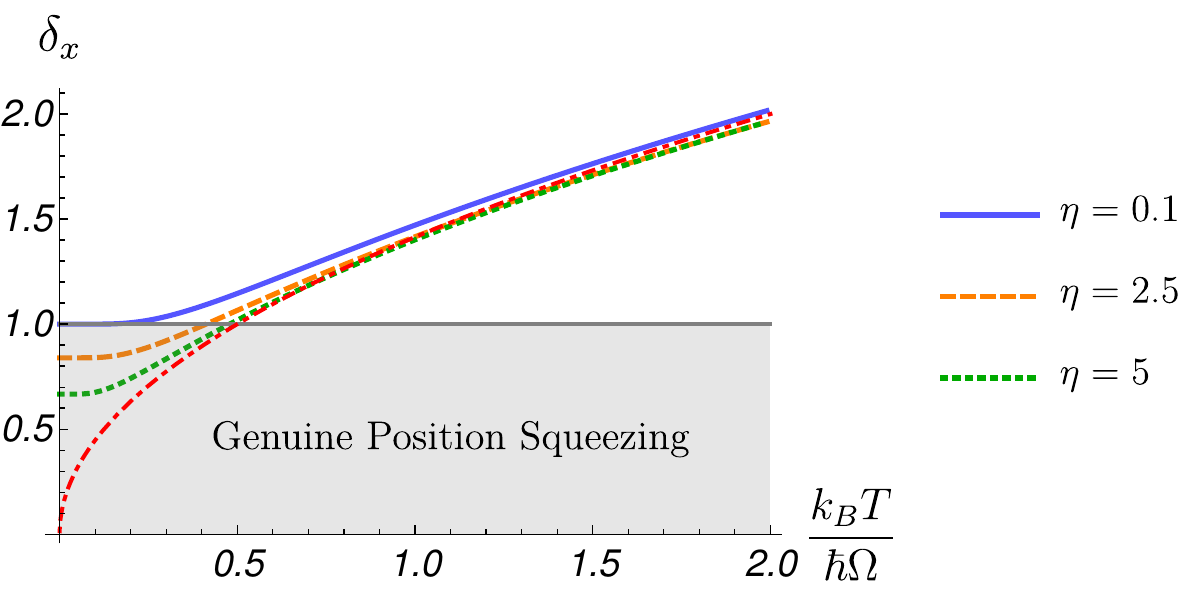}
\caption{\label{PosVarTrap} 
Temperature dependence of the position variance introduced in Eq.\ \eqref{DLessVar}, for different values of the coupling strength.
The plot refers to an impurity of Yb in a trap with a frequency $\Omega=2\pi\cdot200$ Hz, embedded in a Rb gas of $N=5000$ atoms with  trap frequency $\omega_{\rm B}=2\pi\cdot800$ Hz and    coupling strength $g_{\rm B}=0.5\cdot10^{-37}$ J$\cdot$m. 
The red dotdashed line represents the function $\sqrt{2T}$, related to the equipartition theorem. }
\end{center}
\end{figure}
In particular, exploring lower values of the temperature the impurity experiences \textit{genuine position squeezing}, \textit{i.e.} we detect $\delta_x<1$, as shown in Fig.\ \ref{PosVarTrap}. 
The position variance approaches a value smaller than that associated to the Heisenberg principle. 
This implies that, in this regime,  the particle shows less quantum fluctuations in space than in momentum. 
In plain words, the particle is so localized in space, that its position can be measured with an uncertainty which is smaller than that fixed by the Heisenberg principle. 
This effect is enhanced by increasing the value of the coupling strength, while remaining in the regime of low temperatures.  
Note that in the opposite limit, namely at high temperature, the position variance follows the behavior predicted by the equipartition theorem, in agreement with the fact that the uncertainties ellipse approaches the Gibbs-Boltzmann distribution.
We underline that in all the situations we described Heisenberg uncertainty principle is fulfilled at any time and for each values of the system parameters, even when the particle experiences genuine position squeezing. 
This may be checked quickly by evaluating the product between position and momentum variances. 

In comparison with the squeezing predicted for the homogeneous gas, for the inhomogeneous case, one has an extra dependence on the additional parameter, the trapping frequency.  This sets the possibility  of using the BEC trapping frequency to enhance or inhibit the squeezing. 
\begin{figure}
\begin{center}
\includegraphics[width=0.76\columnwidth]{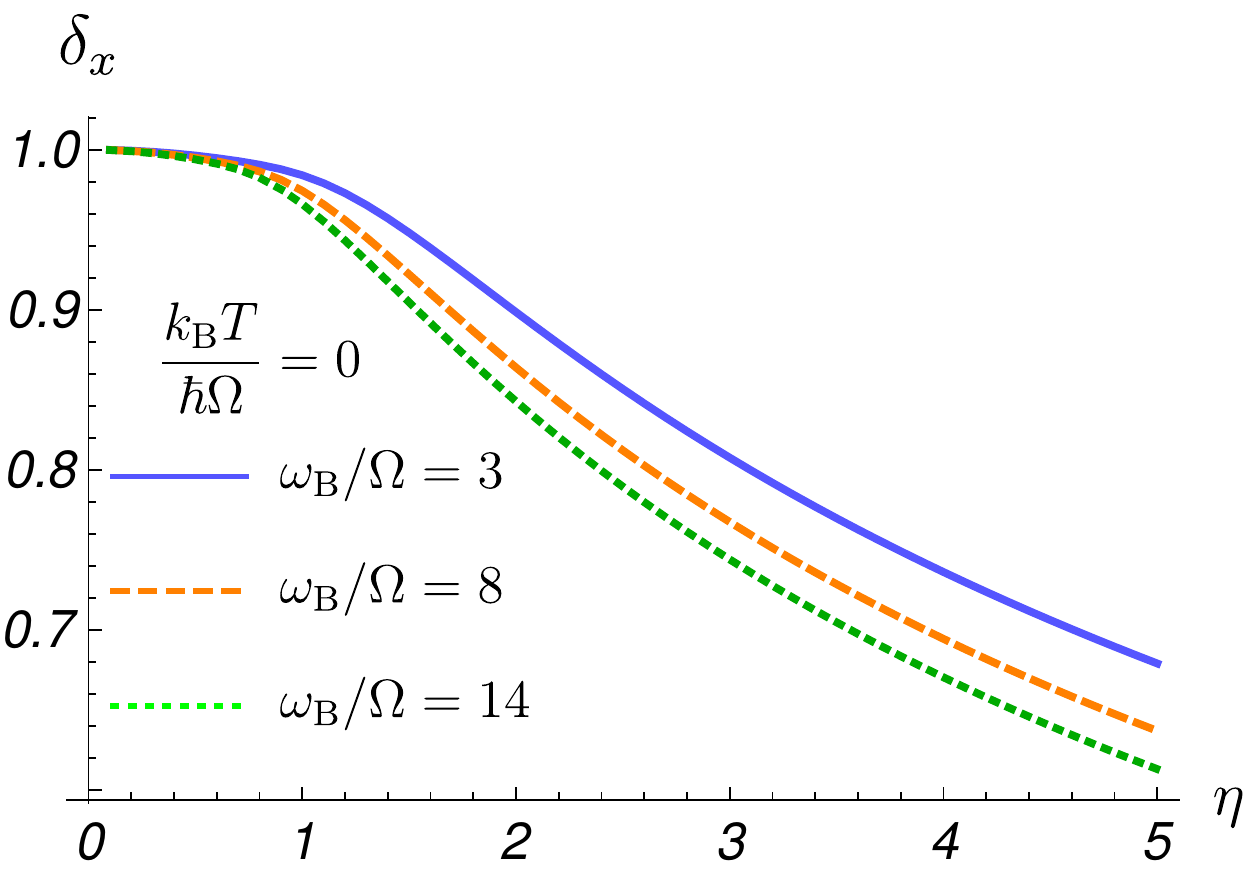}
\caption{\label{PosVarTrapFreq1} 
Position variance introduced in Eq.\ \eqref{DLessVar} as a function of the coupling strength, for different values of the gas trap frequency, in the low temperature regime.
The plot refers to an impurity of Yb in a trap with a frequency $\Omega=2\pi\cdot200$ Hz, embedded in a Rb gas of  $N=5000$ atoms with  trap frequency $\omega_{\rm B}=2\pi\cdot800$ Hz and  coupling strength $g_{\rm B}=0.5\cdot10^{-37}$ J$\cdot$m.  }
\end{center}
\end{figure}
In Fig.\ \ref{PosVarTrapFreq1} we present the position variance as a function of the coupling for several values of the gas trap frequency, in the low-temperature regime.
At weak coupling the gas trap does not play any role and the position variance is approximately equal to one, in agreement with the fact that the impurity approaches the free harmonic oscillator dynamics, collapsing in the ground state ($\delta_x=\delta_p=1$) in the zero-temperature limit. 
As the coupling grows the position variance gets sensitive to the trap of the BEC and we see that genuine position squeezing is enhanced as the BEC trap frequency is made tighter.  
Of course, the dependence on the gas trap frequency is negligible at high-temperature, since in this regime the equilibrium correlation functions get independent on the coupling. 
This may be seen in Fig.\ \ref{posVarTrapFreqT} where we note that, as the temperature grows the position variance approaches a constant value (constant with respect of the frequency) equal to that predicted by the equipartition theorem, in agreement with the behavior presented in Fig.\ \ref{PosVarTrap}. 
\begin{figure}
\begin{center}
\includegraphics[width=0.95\columnwidth]{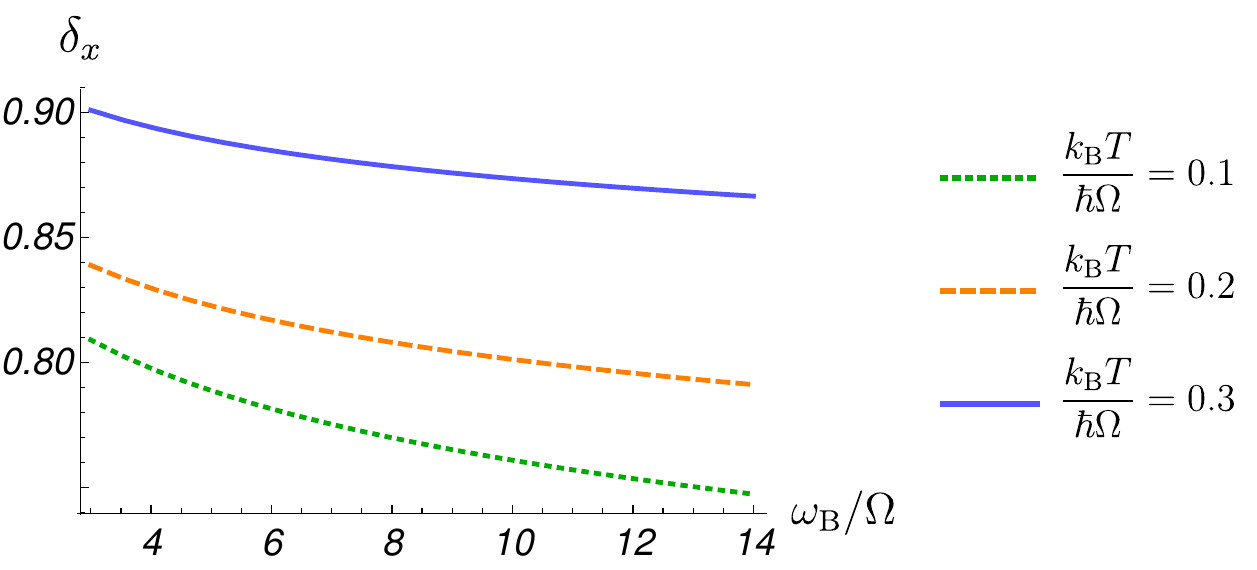}
\caption{\label{posVarTrapFreqT} 
Position variance in Eq.\ \eqref{DLessVar} as a function of the gas trap frequency at several different values of the temperature. 
The plot refers to an impurity of Yb in a trap with a frequency $\Omega=2\pi\cdot200$ Hz, embedded in a Rb gas of $N=5000$ atoms with  trap frequency $\omega_{\rm B}=2\pi\cdot800$ Hz and    coupling strength $g_{\rm B}=0.5\cdot10^{-37}$ J$\cdot$m. }
\end{center}
\end{figure}

In principle one should recover the results obtained for a homogeneous gas by considering the limit in which $\omega_{\rm B}\rightarrow0$. 
This however cannot be seen at the level of the position variance plotted in Figs.\ \ref{PosVarTrapFreq1} and \ref{posVarTrapFreqT}. 
The study of such a limit shows several complications that deserve to be commented. 
We present this discussion in Appendix \ref{LimitApp}. 

Part of the importance of both squeezing and super-diffusion lies in the fact that they may be detected in experiments, since the position variance is a measurable quantity,  
as shown in \cite{Catani2012}. 
Nevertheless, the physical system considered in such an experiment does not fulfill some of the assumptions underlying our theory. 
First of all one has to note that the TF approximation is not satisfied in~\cite{Catani2012}.
A second important difference with the experiment in~\cite{Catani2012} is the initial condition we considered. We assume an initially separated impurity at rest, while in that experimental set-up the laser beam trapping the imputiry gives rise to a different initial condition (see~\cite{Cugliandolo2012}).  

\section{Non-Markovian character of the polaron dynamics}\label{NMSec}
In Sec.\ \ref{HESec} we showed that the inhomogeneous character of the medium alters the analytical form of the SD, and so the dependence on the past history of the system dynamics. 
This is manifested as a different amount of memory effects, namely of the degree of non-Markovianity of the system. 
The purpose of the present section is to evaluate in a quantitative manner the difference of this non-Markovian degree between the cases of a homogeneous and an inhomogeneous gas. 
Note that the study of non-Markovianity in various physical systems and the possibility to tune it by manipulating the related parameters recently attracted a lot of attention, due to the possibility to exploit non-Markovianity as a resource for quantum protocols. 
We quote for instance the important work undertaken in \cite{Liu2011} where a scheme to control non-Markovianity was implemented in an optomechanical-photonic system, and the related, more recent, work in \cite{Haase2018} where the same problem was investigated for an electronic spin diamond. 
In the context of ultracold gases, and in particular of the Bose polaron, an important contribution is represented by the work \cite{Haikka2011}.
Here the authors consider the special case in which the impurity is trapped in a double potential and model such a system by means of the pure-dephasing spin-boson model. 
We treat, instead, the impurity physics in the QBM framework: this is the fundamental difference between our work and \cite{Haikka2011}.

For this goal we select several different techniques, relying on (i) back-flow of information (subsection\ \ref{BFlowInf}); (ii) two-points correlation functions (subsection\ \ref{TwoPointCorrFun}); (iii) ohmic distance (subsection\ \ref{JDist}); (iv) back-flow of energy. 
All these methods show that non-Markovianity is higher when the gas is inhomogeneous. 

\subsection{Back-flow of information}\label{BFlowInf}
We start by quantifying non-Markovianity by means of a measure that associates such a property to the flow of information directed from the environment to the central system, here represented by the impurity, as explained in \cite{Breuer2009}. 
Such an information back-flow may be evaluated taking into account the distinguishability of two initial states: the information coming from the environment allows to better distinguish these states. 
The calculation of this distance is not so complicated for discrete-variable models, while for continuous-variable ones, such as QBM, requires particular attention.
In particular, for the QBM model, the form of the non-Markovianity measure based on back-flow of information has been presented in \cite{Vasile2011}, where it was showed that under particular hypothesis it reads as
\begin{equation}\label{VasileMeasure}
\mathcal{N}_{\rm P}=\int_{\Delta<0}\Delta(t) dt,\quad \Delta(t)=\int^t_0\nu(s)\cos(\Omega s)ds,
\end{equation}
where $\nu(\tau)$ represents the noise kernel in Eq.\ \eqref{NoiseDamp}.
\begin{figure}
\begin{center}
\includegraphics[width=0.95\columnwidth]{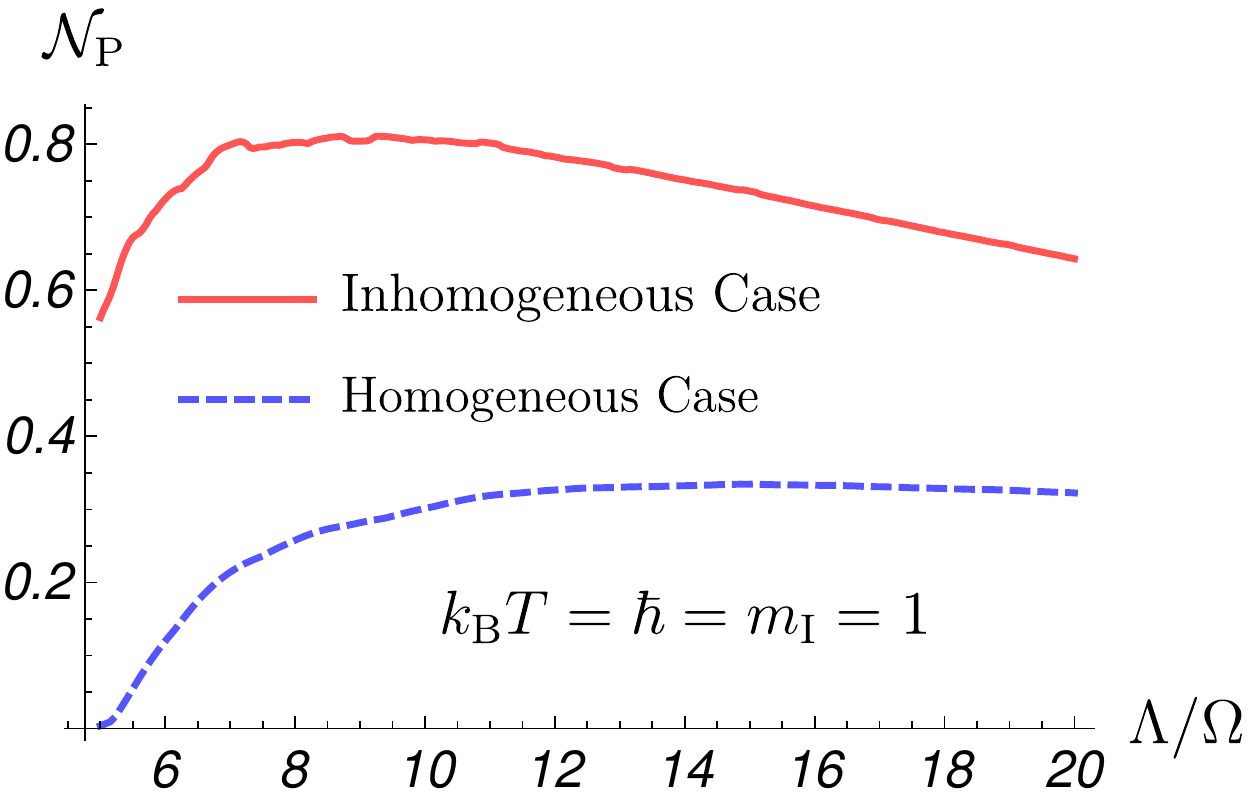}
\caption{\label{plotVasile} 
Non-Markovianity measure in Eq.\ \eqref{VasileMeasure}, as a function of the cut-off frequency, associated to a quartic SD (red-solid line) and a cubic (blue-dashed line) one, corresponding respectively to a inhomogeneous and a homogeneous gas. 
}
\end{center}
\end{figure}

In Fig.\ \ref{plotVasile} we present the measure\ \eqref{VasileMeasure} for the quartic SD in Eq.\ \eqref{spectralDensity} related to an inhomogeneous gas and that derived in \cite{Lampo2017} for a homogeneous medium showing a cubic dependence on the frequency.
Note that in Fig.~\ref{plotVasile} we considered the expression of the noise kernel in the high-temperature regime, namely by approximating the hyperbolic cotangent in Eq.\ \eqref{NoiseDamp} as the inverse of its argument.  The same qualitative behavior is recovered also in the opposite limit, \textit{i.e.} when $T\rightarrow0$ and the cotangent is approximated to one. 
The figure shows that the non-Markovianity degree estimated according the definition in Eq.\eqref{plotVasile} is higher in the inhomogeneous case for any value of the cut-off frequency $\Lambda$. 
Such a result holds for any value of the temperature and the damping constant, since the ratio of the measure computed in the two cases does not depend on these variables. 

\subsection{Two-point correlation function}\label{TwoPointCorrFun}
The result presented in Fig.\ \ref{plotVasile} indicates  that the non-Markovian degree is higher in the inhomogeneous case. 
Nevertheless, one may argue that the measure\ \eqref{VasileMeasure} refers to a map in the pseudo-Lindblad form. 
This is not the case examined in the present manuscript where the polaron dynamics is described by means of Eq.\ \eqref{EqDiffFin}. 
This may be interpreted as a stochastic equation, whose solution is Gaussian and stationary.
A stochastic process is termed \textit{Gaussian} if its joint probability distribution is defined by a normal one. 
In this case, such a feature follows from the fact that the Hamiltonian\ \eqref{analogousQBM} endowed by the interaction term in Eq.\ \eqref{IntHamMiddle} has a quadratic form. 
A process is \textit{stationary} if the joint probability distribution manifests an analytical form that is invariant under temporal translations. 
Such a property can be derived for the present system from the solution in Eq.~\eqref{XHeis}, recalling that also $B(t)$ is stationary. 
Under these hypothesis it has been proven that a stochastic process is Markovian only if it is in the \textit{Ornstein - Uhlenbeck} form, namely its correlation functions decay exponentially in time. 
This statement constitutes a particular form of the \textit{Doob theorem} \cite{MazoBook, BreuerBook}. 
Accordingly, in order to provide a clearcut proof of the non-Markovianity of the system dynamics one has to evaluate the two-point correlation function
\begin{equation}\label{2PCorrFun}
f(t,\tau)\equiv\ave{x(t)x(t+\tau)},\quad \tau>0.
\end{equation}

The quantity in Eq.\ \eqref{2PCorrFun} may be computed starting by the equations of motion in Eqs.\ \eqref{EqXP} and \eqref{Eqbdag} that have to be solved now assuming $t$ as initial time of the system dynamics. 
The solution of the bath modes equation\ \eqref{Eqbdag} takes the form
\begin{align}
b_{k}(t+\tau)&=b_k(t)e^{-i\omega_k\tau}\nonumber\\
&+\frac{g_k}{2}\int^{t+\tau}_tds\exp[+i\omega_k\left(\tau-s\right)]x(s).
\end{align}
Accordingly the equations for the impurity variables get
\begin{align}\label{EqXtau}
\der{x(t+\tau)}{\tau}=\frac{p(t+\tau)}{m_{\rm I}},
\end{align}
and
\begin{align}\label{EqPtau}
\der{p(t+\tau)}{\tau}=&-m_{\rm I}\Omega^2x(t+\tau)+B(\tau,t)\nonumber\\
&-m\der{}{\tau}\int^{t+\tau}_t\Gamma(\tau-s)x(s)ds,
\end{align}
with
\begin{equation}
B(t,\tau)=\sum_{j} \hbar g_{ j}\left[{b}^{\dagger}_{ j}(t)e^{-i\omega_{ j}\tau}+ {b}_{ j}(t)e^{+i\omega_{ j}\tau}\right].
\end{equation}
It is really interesting to note that when the initial condition is translated to a time larger than zero, a dependence on the past-history also enters through the noise term. 
We are interested in the correlation function in Eq.~\eqref{2PCorrFun} so one may proceed by multiplying both sides of Eq.\ \eqref{EqXtau} by $x(t)$ and then taking the average value. 
Thus, deriving both sides with respect of $\tau$ and using Eq.~\eqref{EqPtau} one obtains
\begin{align}
&f_{\tau\tau}(t,\tau)+\Omega^2 f(t,\tau)+\der{}{\tau}\int^{t+\tau}_t\Gamma(\tau-s)f(t,s)ds\nonumber\\
=&\frac{\ave{x(t)B(t,\tau)}}{m_{\rm I}},
\end{align}
where $f_{\tau\tau}$ represents the second-order derivative of $f$ with respect of $\tau$. 
The term in the right-hand side may be treated by recalling Eq.\ \eqref{XHeis}, and assuming that the global bath-impurity state is separable. 
It follows:
\begin{equation}
\ave{x(t)B(t,\tau)}=\frac{1}{m_{\rm I}}\int^{t}_0d\sigma G_{\rm 2}(t-\sigma)\ave{B(0,\sigma)B(t,\tau)}.
\end{equation}
Thus, one can proceed by applying the Laplace transform with respect of the variable $\tau$. It turns:
\begin{align}
&f(t,\tau)=G_{\rm 1}(\tau)\ave{x^2(t)}+G_{\rm 2}(\tau)\ave{x(t)\dot{x}(t)}\nonumber\\+&\frac{1}{m^2_{\rm I}}\int^{t+\tau}_{t}\int^{t}_0dsd\sigma  G_{\rm 2}(t-\sigma)G_{\rm 2}(\tau-s)\ave{B(\sigma)B(s)}\label{2PointCorrFun}.
\end{align}
Note that we take $B(t,\tau)\approx B(0,\tau)\equiv B(\tau)$ because we consider the environment to be large enough in order to assume that its state is constant in time. 
The functions $G_{\rm 1}$ and $G_{\rm 2}$ are those introduced in Eqs.\ \eqref{LTG1} and \eqref{LTG2}, where the variable $z$ is now the frequency associated to $\tau$. 
The average value in the third term of the right hand-side in Eq.\ \eqref{2PointCorrFun} corresponds to 
\begin{equation}
\ave{B(\sigma)B(s)}=\ave{B(\sigma-s)B(0)}=\nu(\sigma-s)-i\eta(\sigma-s), 
\end{equation}
where $\nu(t)$ is the noise kernel\ \eqref{NoiseDamp} and
\begin{align}
\eta(t)=\int^{\infty}_0d\omega J(\omega)\sin\left(\omega t\right)\label{dissipationKernel},
\end{align}  
is the dissipation kernel. 

The expressions of the noise and damping kernel, together with those of $G_{\rm 1}$ and $G_{\rm 2}$ determine the analytical structure of the two-point correlation function. 
To obtain the final expression of this, one needs the explicit form of $G_{\rm 1}$ and $G_{\rm 2}$ and so has to invert the Laplace transforms in Eqs.\ \eqref{LTG1} and \eqref{LTG2}. 
Such a problem has already been treated in Sec.\ \ref{PosVarSec} for both a trapped ($\Omega>0$) and untrapped ($\Omega=0$) impurity.
In the first situation it has been shown that the Laplace transforms have to be inverted numerically. 
In this manner, anyway, it is not possible to derive an explicit expression for them. 
To reduce such a problem to an analytically feasible one, we can expand the Laplace transform of the damping kernel appearing in the denominators of Eqs.\ \eqref{LTG1} and \eqref{LTG2} to the first order in $z/\Lambda$, obtaining two expressions that may be inverted analytically. 
 This gives the following form for the Green functions,  
\begin{equation}\label{G12TrapApp}
G_{\rm 1}(t)=\cos\left(\frac{\Omega}{\sqrt{\tilde{\alpha}}} t\right),\quad G_{\rm 2}(t)=\frac{1}{\sqrt{\tilde{\alpha}}\Omega}\sin\left(\frac{\Omega}{\sqrt{\tilde{\alpha}}} t\right).
\end{equation}
The oscillating functions above do not reproduce the behavior presented in Fig.\ \ref{G1-2Trap}. 
The exact temporal dependence of $G_{\rm 1}$ and $G_{\rm 2}$ obtained by means of the Zakian numerical method shows at very long time a damping and a time-dependent renormalization of the frequency. 
So, the regime of validity of the result in Eq.\ \eqref{G12TrapApp} has to be discussed carefully.
These expressions have been obtained by considering an expansion in $z/\Lambda$ at the first order and thus they describe a long-time regime that quantitatively means $\Lambda t\gg1$. 
Note that Fig.\ \ref{G1-2Trap} refers to $\Lambda/\Omega=\omega_{\rm B}/\Omega=4$, accordingly any time $\Omega t\gg\Omega/\Lambda=0.25$, for instance $\Omega t=10\Omega/\Lambda=2.5$, maybe considered as a "long" one in such a specific situation.
Here, it is possible to check that the functions\ \eqref{G12TrapApp} match the oscillating non-damped behavior in Fig.\ \ref{G1-2Trap} for $\Omega t\lesssim 20$.

The oscillating behavior in Eq.\ \eqref{G12TrapApp} would be enough to state that even in presence of a trap the system dynamics is non-Markovian since no exponential decays occur. 
One could try to compute the whole correlation function for the sake of completeness, but the approximated expressions in Eq.\ \eqref{G12TrapApp} do not ensure the convergence of the integrals in the third term in the right hand-side of Eq.\ \eqref{2PointCorrFun}. 
Of course, this is an unphysical effect which vanishes if one considers more accurate expressions for $G_{\rm 1}$ and $G_{\rm 2}$ that include also the damping. 
One should expand so the Laplace transform of the damping kernel beyond the first-order, but this leads to a logarithmic dependence on $z$ that forbids the inversion of the Laplace transforms in an analytical manner.

\begin{figure}
\begin{center}
\includegraphics[width=0.95\columnwidth]{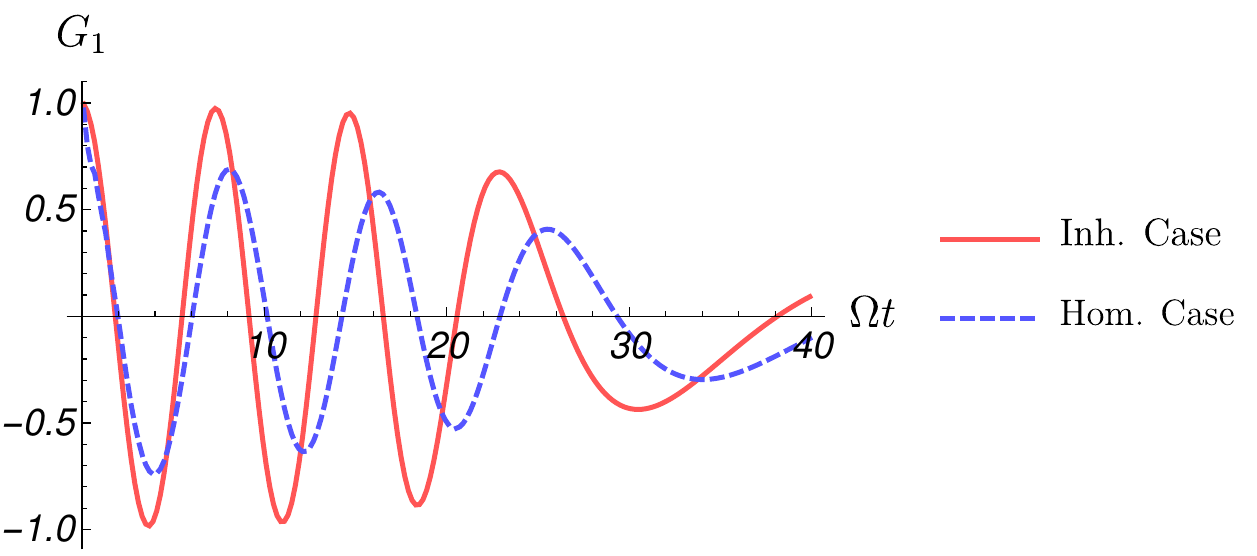}
\includegraphics[width=0.95\columnwidth]{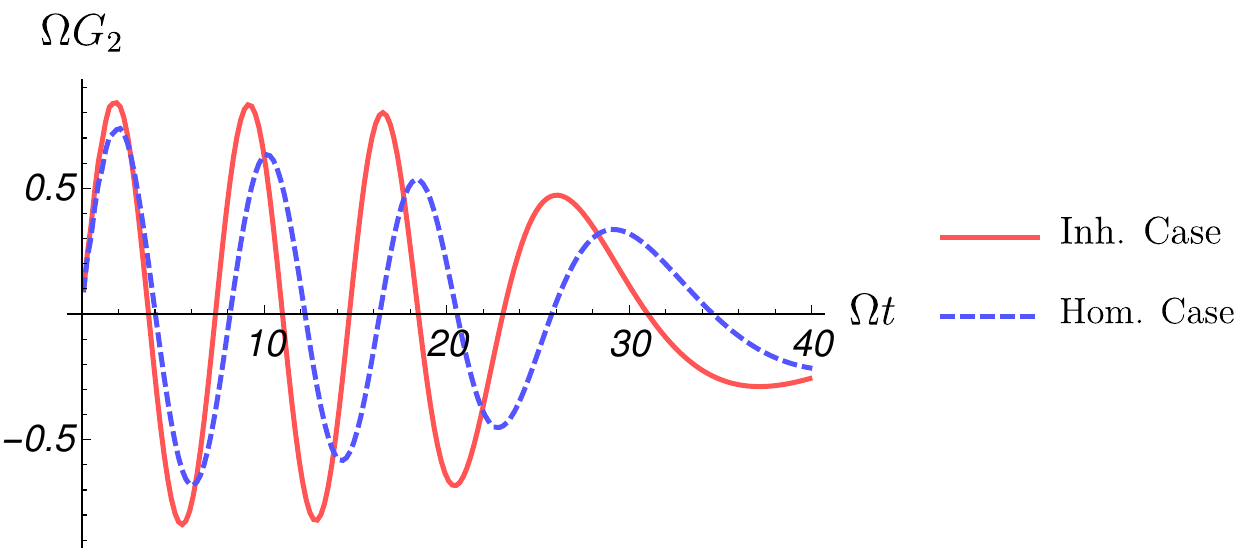}
\caption{\label{G1-2TrapComp} Time dependence of $G_{\rm 1}$ (up) and $G_{\rm 2}$ (down) calculated for the quartic SD in Eq.\ \eqref{spectralDensity} related to a inhomogeneous gas (solid red line) and a cubic one derived in \cite{Lampo2017} for a homogeneous medium (dashed blue line). 
The plot has been realized for $\Lambda/\Omega=10$ and $\gamma/\Omega=7$.
}
\end{center}
\end{figure}

The first two terms in the right hand-side in Eq.\ \eqref{2PointCorrFun} play an important role in the analysis of the memory effect because they rule the decay of the initial position and velocity. 
We can study its form in the homogeneous and inhomogeneous case in order to establish in which situation the non-Markovian degree is higher. 
The approximated expressions\ \eqref{G12TrapApp} are not suitable for this task, thus we compare the exact numerical result, as shown in Fig.\ \ref{G1-2TrapComp}.  
Here it is possible to see that both $G_{\rm 1}$ and $G_{\rm 2}$ calculated in the homogeneous case decay faster than those obtained in the inhomogeneous one. 
This suggests that the effect of the past history on the system dynamics vanishes faster if the medium is inhomogeneous. 

We treat now the same problem in the case where $\Omega=0$ (untrapped impurity). 
\begin{figure}
\begin{center}
\includegraphics[width=0.95\columnwidth]{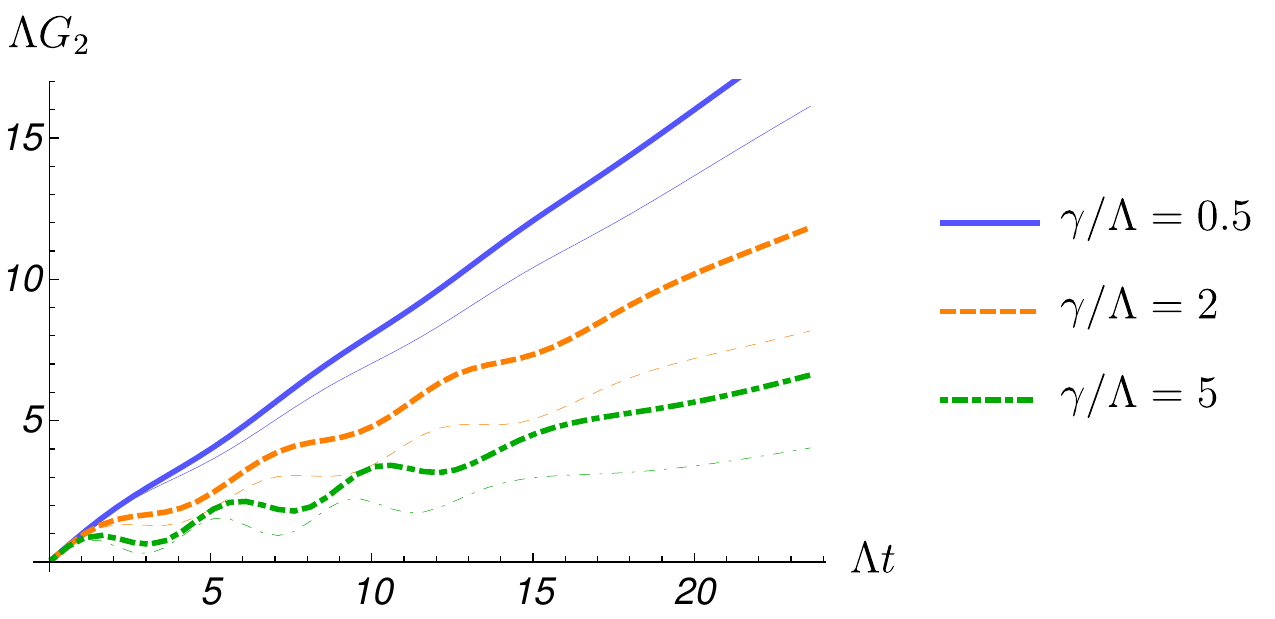}
\caption{\label{G2UntrapComp} Time-dependence of the function $G_{2}$, defined through its Laplace transform in Eq.\ \eqref{LTG2}. 
The thick lines represent the result obtained in the inhomogeneous case, while the thin ones refer the homogeneous medium. 
}
\end{center}
\end{figure}
In this case the function $G_{\rm 1}$ is identically equal to $1$, while $G_{\rm 2}$ shows the ballistic form presented in Eq.\ \eqref{G2An}. 
Such a ballistic behavior is enough to state that even when $\Omega=0$ we recover a non-Markovian dynamics since no exponential decays occur.
Still, we can compare the form of $G_{\rm 2}$ in the homogeneous and inhomogeneous case to establish which dynamics is "less Markovian". 
In Fig.\ \ref{G2UntrapComp} we see that for each value of the damping constant $\gamma$, and at any time, the value of $G_2$ is higher in the inhomogeneous case. 
This means that the dependence on the initial condition, \textit{i.e.} the past history of the system, is stronger and thus we find again that the inhomogeneous case is the "less Markovian". 

The situation in which the impurity is untrapped is very interesting because one may exploit the long-time analytical expression for $G_{\rm 2}$ in Eq.\ \eqref{G2An} to derive the whole correlation expression.
This, at the best of our knowledge, constitutes an original calculation. 
For this goal one may decompose the hyperbolic cotangent appearing in the noise kernel as a sum over the Matsubara frequencies $\nu_n=2\pi k_{\rm B}T n/\hbar$,
\begin{equation}
\coth\left(\frac{\hbar\omega}{2k_{\rm B}T}\right)=\frac{2k_{\rm B}T}{\hbar}\left(\frac{1}{\omega}+2\sum^\infty_{n=1}\frac{\omega}{\omega^2+\nu^2_n}\right).
\end{equation}
Replacing this expression into Eq.\ \eqref{NoiseDamp} one gets
\begin{equation}
\nu(t)=\nu^{\rm{\rm{(HT)}}}(t)+\nu^{\rm{(LT)}}(t),
\end{equation}
with
\begin{align}
&\nu^{\rm{\rm{(HT)}}}(t)=\frac{2k_{\rm B}T}{\hbar}\int^{\infty}_0d\omega \frac{J(\omega)}{\omega}\cos\left(\omega t\right),\\
&\nu^{\rm{(LT)}}(t)=\frac{2k_{\rm B}T}{\hbar}\sum^{\infty}_{n=1}\int^{\infty}_0d\omega J(\omega)\frac{\omega}{\omega^2+\nu^2_n}\cos\left(\omega t\right)\label{nuLT},
\end{align}
ruling respectively the high-temperature regime and the low-temperature one. 
Therefore, recalling Eq.\ \eqref{G2An}, the two-point correlation function\ \eqref{2PCorrFun} takes the form
\begin{align}
\ave{x(t)x(t+\tau)}=&\ave{x^2(t)}+\frac{\tau}{\tilde{\alpha}}\ave{x(t)\dot{x}(t)}\nonumber\\+&\mathcal{I}_{\eta}+\mathcal{I}^{\rm{(HT)}}_{\nu}+\mathcal{I}^{\rm{(LT)}}_{\nu}\label{2PointCorrFun2},
\end{align}
in which
\begin{align}
\mathcal{I}_{\eta}&=\frac{1}{m^2_{\rm I}\tilde{\alpha}}\int^\infty_0d\omega J(\omega)\int^t_0d\sigma(t-\sigma)\nonumber\\&\times\int^{t+\tau}_tds(\tau-s)\sin\left[\omega(\sigma-s)\right],
\end{align}
and 
\begin{align}
&\mathcal{I}^{\rm{(HT)}}_{\nu}=\frac{2k_{\rm B}T}{\hbar m^2_{\rm I}\tilde{\alpha}}\int^\infty_0d\omega \frac{J(\omega)}{\omega}\tilde{\mathcal{I}}_{\nu}(\omega),\\
&\mathcal{I}^{\rm{(LT)}}_{\nu}=\frac{2k_{\rm B}T}{\hbar m^2_{\rm I}\tilde{\alpha}}\sum^\infty_{n=1}\int^\infty_0d\omega \frac{J(\omega)\omega}{\omega^2+\nu^2_n}\tilde{\mathcal{I}}_{\nu}(\omega),\label{InuLT}
\end{align}
with
\begin{align}
\tilde{\mathcal{I}}_{\nu}(\omega)=\int^t_0d\sigma(t-\sigma)\int^{t+\tau}_tds(\tau-s)\cos\left[\omega(\sigma-s)\right].
\end{align}
In particular we will focus on the situation in which $T=0$. 
In this case Eq.\ \eqref{InuLT} takes the form
\begin{equation}
\mathcal{I}^{(0)}_{\nu}=\frac{1}{m^2_{\rm I}\tilde{\alpha}}\int^{\infty}_0d\omega J(\omega)\tilde{\mathcal{I}}_{\nu}(\omega).
\end{equation}
Although the ballistic form of $G_2$ would be enough to prove the fact that the correlation function does not decay exponentially, we derive for sake of completeness the expression of all the terms. 
This, to the best of our knowledge, has never been investigated before for the present case. 

In the long-time limit we have
\begin{align}
\tilde{\mathcal{I}}_{\eta}(\omega)=&\frac{\gamma t}{2m_{\rm I}\tilde{\alpha}^2\Lambda^3}\left[\Lambda^2+\frac{2}{\tau^2}+2\Lambda^2\cos\left(\Lambda t\right)\right]\nonumber\\
+&\frac{\gamma t^4\Lambda^2}{m_{\rm I}\tilde{\alpha}^2}\frac{\cos\left[\Lambda(t+\tau)\right]}{\Lambda^3\left(t+\tau\right)^3}\nonumber\\
+&\frac{\gamma t}{m_{\rm I}\tilde{\alpha}^2\Lambda^3\tau^2}\left[\cos\left(\Lambda\tau\right)+\Lambda\tau\sin\left(\Lambda\tau\right)\right],
\end{align}
\begin{align}
\tilde{\mathcal{I}}^{\left(HT\right)}_{\nu}(\omega)=&\frac{2k_{\rm B}T\gamma t}{m_{\rm I}\tilde{\alpha}^2\Lambda^3\tau\left(t+\tau\right)^2}\left[\cos\left(\Lambda t\right)-1\right]\nonumber\\
+&\frac{2k_{\rm B}T\gamma t\Lambda\tau}{m_{\rm I}\tilde{\alpha}^2\Lambda^3\tau\left(t+\tau\right)^2}\left[\sin\left[\left(t+\tau\right)\right]-\sin\left(\Lambda t\right)\right],
\end{align}
\begin{align}
\tilde{\mathcal{I}}^{\left(0\right)}_{\nu}(\omega)=
&\frac{\gamma t}{\Lambda^3}\left[\frac{\sin\left(\Lambda\tau\right)}{\tau^2}-\frac{\Lambda}{\tau}\cos\left(\Lambda\tau\right)\right]\nonumber\\
+&\frac{\gamma t}{\Lambda}\left[\frac{t^3\sin\left[\Lambda\left(t+\tau\right)\right]}{\left(t+\tau\right)^3}-\sin\left(\Lambda t\right)\right].
\end{align}
The equations above show a ballistic dependence on time, in agreement with the fact that the impurity is untrapped. 
This particular temporal behavior definitely proves that the correlation function does not decay exponentially and, in the end, the process is not Markovian. 
Note that, in principle, one should recover the expressions for the MSD derived in Sec.\ \ref{NoImpTrapSec} by taking the limit in which $t\rightarrow0$.
This does not follow by the equations above because they refer to the long-time limit, \textit{i.e.} $\Lambda t\gg1$. 

\subsection{J-Distance}\label{JDist}
In order to study in detail the comparison between the amount of memory effects occuring in an inhomogeneous and a homogeneous gas we introduce a quantifier strictly related to the class of equations with the form showed in Eq.\ \eqref{EqDiffFin}:
\begin{equation}\label{OhmMeas}
\mathcal{N}^{\rm (J)}=\md{\frac{\ave{x^2_{\rm J}}-\ave{x^2_{\rm Ohm}}}{\ave{x^2_{\rm J}}+\ave{x^2_{\rm Ohm}}}},
\end{equation}
where $\ave{x^2_{\rm J}}$ and $\ave{x^2_{\rm Ohm}}$ constitute the position variance calculated respectively with a given SD, $J$, and the ohmic one, \textit{i.e.} that exhibiting a linear dependence on frequency in the limit in which such a variable is much smaller than $\Lambda$. 
It is very important to point out that the quantity in Eq.\ \eqref{OhmMeas} does not measure the distance from \textit{a generic} Markovian process, but from \textit{a particular} one, given by the Langevin equation\ \eqref{EqDiffFin} with an ohmic spectral density. 
Nevertheless one has to note, recalling Eq.\ \eqref{DampingKernel}, that the only form of the SD leading to a completely local-in-time Langevin equation (resulting from a Dirac delta damping kernel) is the ohmic one.
Then, the measure in Eq.\ \eqref{OhmMeas} quantifies the difference between the position variance calculated for the present system and that obtained by means of the Markovian form of Eq.\ \eqref{EqDiffFin}: when $\mathcal{N}$ tends to zero the distance from such a Markovian process is minimum, while it is maximum when $\mathcal{N}$ is close to one. 
Of course, because of its definition, the measure does not take any value outside $[0,1]$.
\begin{figure}
\begin{center}
\includegraphics[width=0.95\columnwidth]{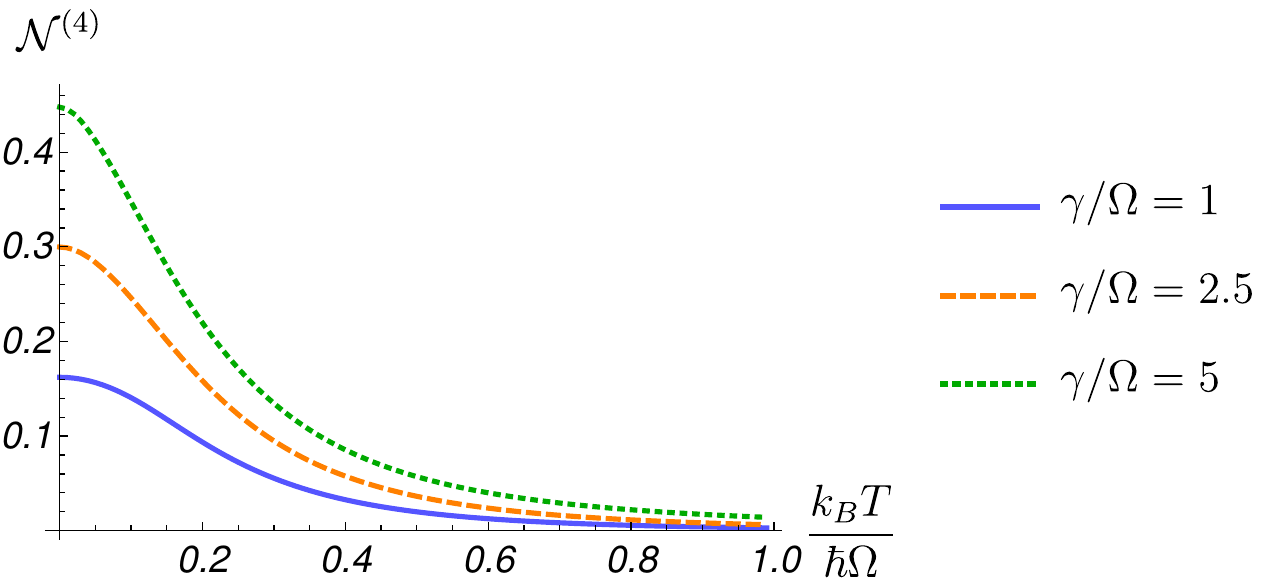}
\caption{\label{N4Plot} 
Non-Markovianity measure in Eq.\ \eqref{OhmMeas}, associated to SD in Eq.\ \eqref{spectralDensity}, as a function of the temperature for different values of the damping constant. 
The plot refers to $\Lambda/\Omega=10$. }
\end{center}
\end{figure}

The quantity in  Eq.\ \eqref{OhmMeas} is shown in Fig.\ \ref{N4Plot}.
We point out that the difference with the Markovian ohmic process grows in the zero-temperature limit, while vanishes as the temperature increases. 
This is in agreement with the fact that in the high-temperature regime the particle approaches a Gibbs-Boltzmann state and its variances follows the behavior predicted by the equipartition theorem, as shown in Fig.\ \ref{PosVarTrap}, \textit{i.e.} they do not dependent on the coupling and thus on the SD. 
Accordingly the difference between two position variances computed with any pair of different SD tends to zero. 
We also note that $\mathcal{N}^{(4)}$ vanishes as the damping constant decreases, in agreement with the fact that when this parameter goes to zero, the physics of the system gets coupling independent. 
Finally, we note that the dynamics of an impurity in a trapped BEC approaches that of a Markovian system at high temperature and weak coupling. 

\begin{figure}
\begin{center}
\includegraphics[width=0.95\columnwidth]{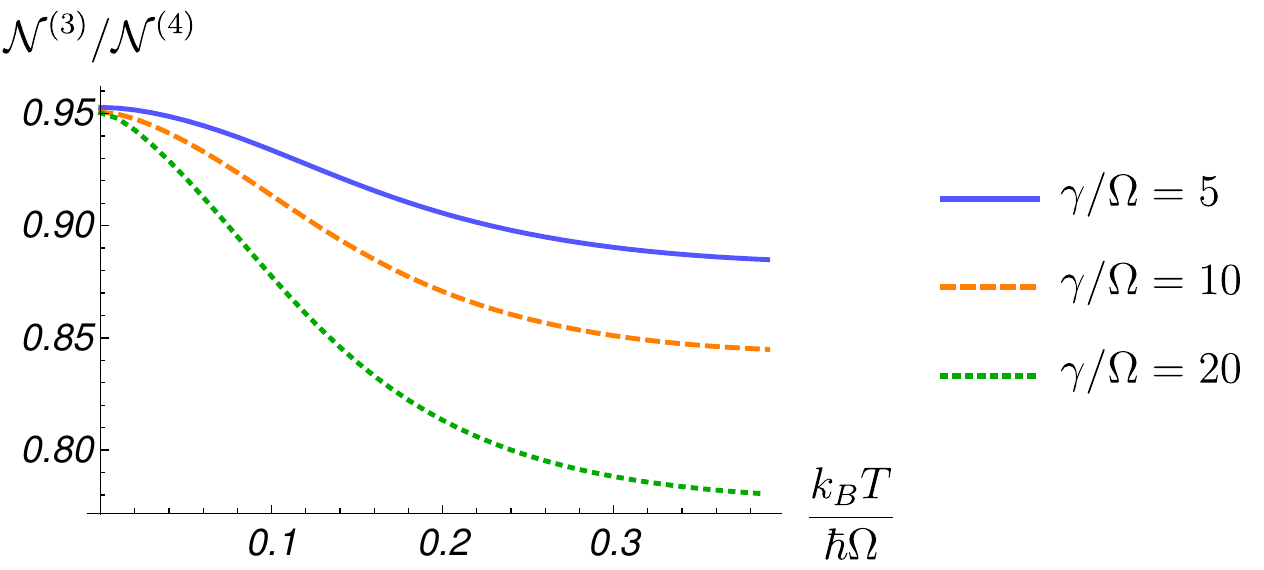}
\caption{\label{CompPlot} 
Ratio between the non-Markovianity measure in Eq.\ \eqref{OhmMeas} calculated for the SD associated to the homogeneous case (see Eq.\ (41) in \cite{Lampo2017}) and that in Eq.\ \eqref{spectralDensity}. 
The plot expresses a temperature dependence, for different values of the damping constant and refers to $\Lambda/\Omega=10$. }
\end{center}
\end{figure}
In Fig.\ \ref{CompPlot} we aim to compare the value of the measure for the inhomogeneous case with that of the homogeneous one, at a given value of the temperature and damping constant. 
We see that the distance is higher for the former, and the difference grows at low temperature and as the coupling increases. 

\subsection{Back-flow of energy}
We conclude the discussion concerning the non-Markovian degree of the polaron dynamics by considering a further criterion based on the back-flow of energy. 
In \cite{Guarnieri2016} it has been shown that there is a correlation between the non-Markovian character of the dynamics and the emergence of a back-flow of energy, namely a flow of energy directed from the environment to the central system. 
The evaluation of the back-flow of energy for the super-ohmic SDs model has, at the best of our knowledge, never been explored. 
This is the purpose of the present subsection.
We evaluate therefore 
\begin{equation}\label{BFlowEnergy}
\Phi_{\rm \epsilon}=\int_{\partial_tE>0}\der{E(t)}{t}dt,\quad E(t)=\frac{\ave{p^2(t)}}{2m_{\rm I}}, 
\end{equation}
where the expression for the impurity momentum can be obtained by deriving the position operator in the Heisenberg picture in Eq.\ \eqref{XHeis} with respect to time.
We perform this calculation in the case in which the impurity is untrapped, since we may exploit the long-time analytical expression for $G_{\rm 2}$ in Eq.\ \eqref{G2An}. 
In addition, in the context of the energy back-flow analysis the untrapped case is more interesting because allows to get rid of the energy flux due to the oscillations related to the impurity trap and permits to focus only on those associated to the interaction with the bath. 
\begin{figure}
\begin{center}
\includegraphics[width=0.95\columnwidth]{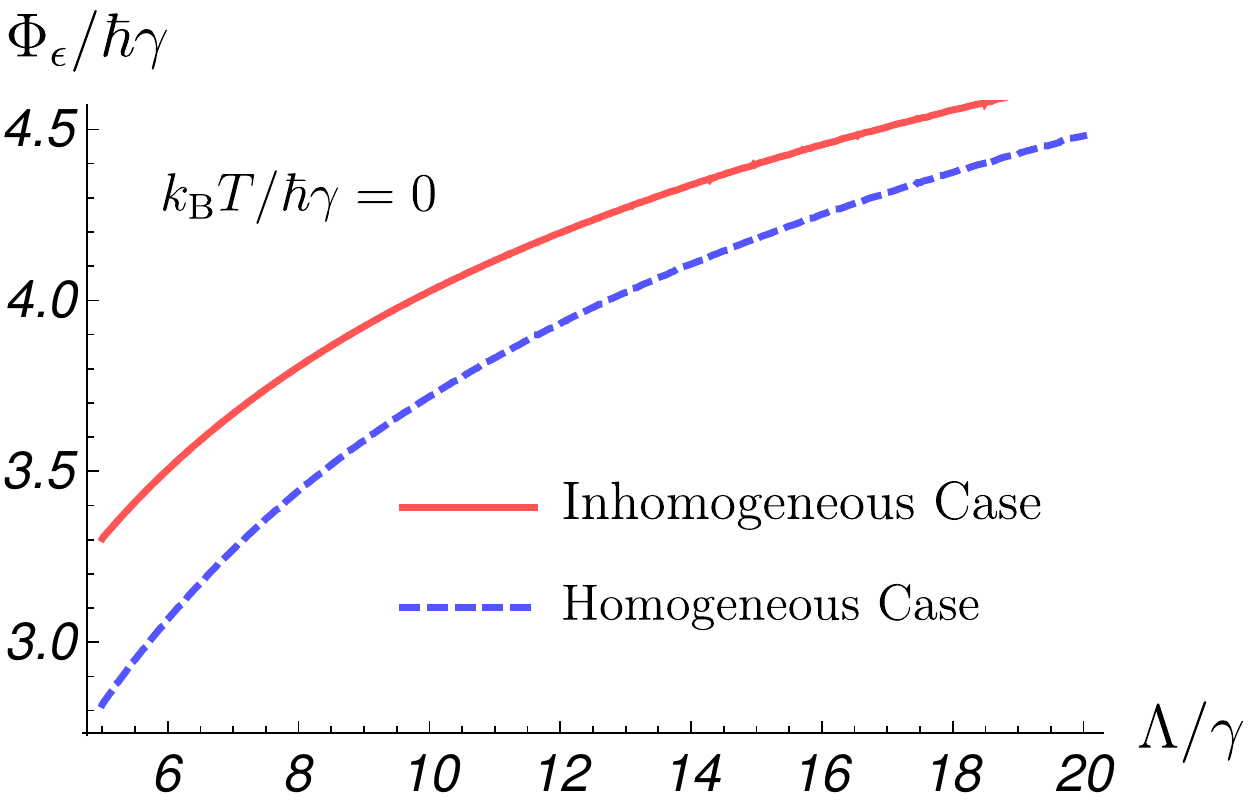}
\caption{\label{BackFlowPlot} 
Back-flow of energy in Eq.\ \eqref{BFlowEnergy} as a function of the cut-off frequency calculated for a homogeneous gas and an inhomogeneous gas (red solid line) and a homogeneous one (blue dashed line). }
\end{center}
\end{figure}
In Fig.\ \ref{BackFlowPlot} we plotted the quantity in Eq.\ \eqref{BFlowEnergy} in both the homogeneous and inhomogeneous case. 
It shows that the flow of energy coming from an inhomogeneous environment is always larger than that coming form a homogeneous one. 
The picture is plotted for the low-temperature regime but we find the same qualitative behavior in the opposite limit. 
It is also interesting to note that,  both in the homogeneous and inhomogeneous case, $\Phi_{\rm \epsilon}$ grows as the cut-off frequency increases. 
This admits a microscopic interpretation: when the cut-off frequency increases the number of bath modes coupled to the impurity grows, so the flux of energy is bigger. 

\section{Conclusions and perspectives}
We presented a study of the dynamics of an impurity in an inhomogeneous Bose-Einstein condensate. 
Such a problem is treated in the framework of open quantum systems, as it can be brought formally to the form of the quantum Brownian motion model. 
The main motivation to do this lies in the possibility to analyze in detail the out-of-equilibrium dynamics of the impurity. 
The inhomogeneous character of the BEC, due to the presence of an external confining trap, strongly modifies the properties of the impurity-bath coupling.
In general, such an interaction shows a non-linear dependence on the position of the central particle. 
One could treat the corresponding dynamics by recalling the theory developed in \cite{Barik2005}, where the Heisenberg equations for the QBM with a non-linear coupling have been derived. 
Nevertheless, these results cannot be applied straightforwardly, since in the present case, we have a different analytical dependence on the position for each value of $k$. 
We approximate thus this interaction by a linear function, provided that the analysis is restricted to the middle of the trap. Under this assumption, one reproduces formally the situation of the traditional quantum Brownian motion model. 
This approximation results to be totally appropriate for the regime parameters we considered, as discussed in Appendix\ \ref{ValLinApp}. 

We derive the Langevin equation for the impurity position in the Heisenberg picture and we calculate the spectral density. 
Here we detect an important difference with the study presented in \cite{Lampo2017} for a homogeneous gas: the inhomogeneity of the medium results to a higher super-Ohmic degree, suggesting that the amount of memory effects carried out by the impurity is bigger.

Such an issue has been treated in a quantitative manner in Sec.\ \ref{NMSec}.
We employed four different criteria to evaluate non-Markovianity and all of these indicated that the amount of memory effects increases when the gas is confined in a trap.
The higher non-Markovianity degree for an inhomogeneous medium represents the main qualitative change with respect to the homogeneous case studied in \cite{Lampo2017}.
Non-Markovianity attracted a lot of interest during the last years \cite{Liu2011, Haikka2011, Guarnieri2016, Tudela2017, Strasberg2018} especially in view of the possibility to exploit it as a resource for quantum devices. 
For instance in \cite{Vasile2011b} it has been proved that quantum key distribution protocols in non-Markovian channels provide alternative ways of protecting the communication which cannot be implemented in usual Markovian channels. 

Nevertheless, the results we presented just constitute a first step for a quantitative analysis for the control of memory effects in polaron dynamics.
For this goal there are also other techniques that one could recall, such as that in~\cite{Vasile2014}, where the effect of the cut-off in the memory effects is elucidated. In our  comparison between the memory effects  in the homogeneous and inhomogeneous BEC we focused in the degree of the superohmicity of the spectral density. A  study on the effect of the cut-off is interesting but falls beyond the scope of the present paper.


If we embed the impurity particle in a harmonic potential the position and momentum variances in the long-time limit reach a stationary value. That is, the particle reaches equilibrium in the long-time limit, with quantum fluctuations independent of time. 
We study  its behaviour once this equilibrium is reached as a function of the parameters that may be tuned in experiments, such as temperature and gas-impurity coupling strength. 
At low-temperatures and by increasing the value of the coupling we find that the particle experiences genuine position squeezing, \textit{i.e.} $\delta_x<1$. This corresponds to high-spatial localization, i.e., the quantum fluctuations in space are smaller than those in momentum in terms of the uncertainty ellipse. 
Very importantly, we show that the spatial squeezing can be controlled with the BEC trap frequency, particularly it is enhanced as this frequency is increased.   
Genuine position squeezing can be detected in experiments, as the position variance represents a measurable quantity. The fact that the squeezing can be controlled with the BEC trap frequency has important implications for the verification of these effects in current experiments. 

In general, the application of the quantum Brownian motion to this realistic system opens the possibility to look in the concrete case of Bose polaron for the large number of effects detected at an abstract level for the general model. 
For instance, one could try to propose an experiment with ultra-cold gases to study the Zeno effect predicted in \cite{Maniscalco2006}. 
Moreover, it is possible to study in the context of the Bose polaron the emergence of classical objectivity, that has been study for open quantum systems in \cite{Tuziemski2015, Lampo2017b}. 

\acknowledgments

Insightful discussion with Philipp Strasberg, Jan Wehr, Roberta Zambrini, Jacopo Catani and Giulia de Rosi are  gratefully  acknowledged. 
This work has been funded by a scholarship from the Programa M\`{a}sters d'Excel-l\`{e}ncia of the Fundaci\'{o} Catalunya-La Pedrera, ERC Advanced Grant OSYRIS, EU IP SIQS, EU PRO QUIC, 
EU STREP EQuaM (FP7/2007-2013, No. 323714).
M. L. acknowledges the Spanish Ministry MINECO (National Plan
15 Grant: FISICATEAMO No. FIS2016-79508- P, SEVERO OCHOA
No. SEV-2015- 0522), Fundació Cellex, Generalitat de Catalunya
(AGAUR Grant No. 2017 SGR 1341 and CERCA/Program), ERC
AdG OSYRIS, EU FETPRO QUIC, and the National Science
Centre.

\appendix

\section{Validity of the linear approximation for the dynamics in the middle of the gas trap}\label{ValLinApp}
The results presented for both a trapped and an untrapped impurity have been derived by approximating the interaction Hamiltonian in Eq.\ (\ref{generalIntHam}) as a linear function of the position impurity. 
Such a linear expansion is valid in the middle of the trap, i.e. when 
\begin{equation}
x\ll R \label{valCond}
\end{equation}
In this part, we study the validity of the condition\ (\ref{valCond}) as the parameters of the system vary. 
For this goal we distinguish the situation where the impurity is trapped ($\Omega>0$) and that in which it is untrapped ($\Omega=0$). 

For the trapped impurity, in general, the condition in Eq.\ (\ref{valCond}) may be expressed as
\begin{equation}\label{valCond2}
x\approx\ave{x}+\delta_{x}=\Delta_{ x}\ll R, 
\end{equation}
where $\Delta_x$ is the Gaussian deviation of the position from its average value. 
At low temperatures such a condition is usually fulfilled because the position variance of the impurity achieves very low values, since the particle experiences squeezing.
In order to evaluate Eq.\ \eqref{valCond2} we recall the values acquired by the dimensionless variance $\delta_x$. 
For instance, for the system parameters used in Fig.\ \ref{PosVarTrap}, it turns
\begin{equation}
\delta_x\ll (R/a_{\rm HO})\lesssim 11 ,
\end{equation}
where $a_{\rm HO}=\sqrt{\hbar/m_{\rm I}\Omega}$ is the impurity harmonic oscillator length. 

At high temperatures instead, the position variance approaches the behavior predicted by the equipartition theorem, \textit{i.e.}
\begin{equation}
\delta_{x}\approx\sqrt{\frac{2k_{\rm B}T}{m_{\rm I}\Omega^2}}.
\end{equation}
Accordingly, the condition in Eq.\ (\ref{valCond2}) induces maximum acceptable temperature 
\begin{equation}
T_{\rm crit}=m_{\rm I}\Omega^2R^2/k_{\rm B}.
\end{equation}
In particular, for the values of the physical quantities employed in Fig.\ \ref{PosVarTrap}
\begin{equation}
\frac{k_{\rm B}T_{\rm crit}}{m_{\rm I}\Omega^2a^2_{\rm HO}}\lesssim 122. 
\end{equation}

We now study the validity condition in Eq.\ (\ref{valCond2}) for an untrapped impurity. In this case it may be expressed as
\begin{equation}\label{valCondUntrapped}
\text{MSD}(t)\ll R^2,
\end{equation}
inducing a constraint on the time and on the interaction strength. 
Precisely, replacing Eq.\ (\ref{MSDLT}) in Eq.\ (\ref{valCondUntrapped}), we obtain, in the particular case in which $\ave{\dot{x}^2(0)}=0$, that the linear approximation when $\Omega=0$ is valid provided
\begin{equation}\label{validUntrap}
\frac{1}{3\tilde{\alpha}^2}\left(\frac{\hbar\gamma(\eta)}{m_{\rm I}}\right)\left(\frac{t}{R}\right)^2\ll1. 
\end{equation}
\begin{figure}
\begin{center}
\includegraphics[width=0.95\columnwidth]{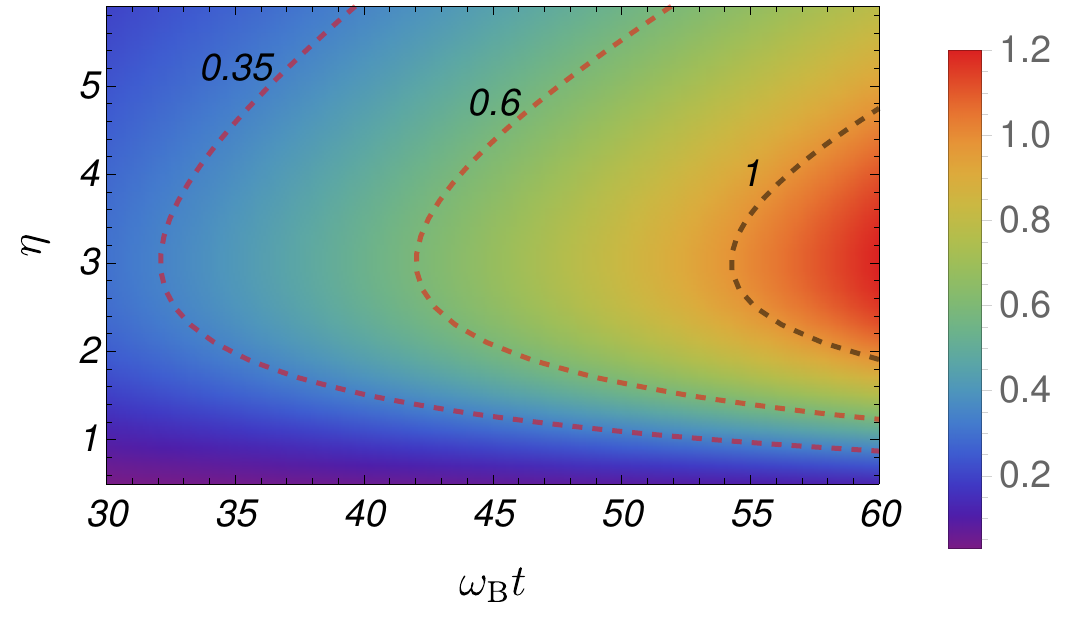}
\caption{\label{validUntrapFig}Validity condition in Eq.\ (\ref{validUntrap}) for an untrapped impurity of Yb in a gas made up by $N=5000$ atoms of K with a coupling strength $g_{\rm B}=0.5\cdot10^{-37}$J$\cdot$m, trapped in a harmonic potential with $\omega_{\rm B}=2\pi\cdot800$ Hz.
}
\end{center}
\end{figure}
The left hand-side of Eq.\ (\ref{validUntrap}) is plotted in Fig.\ \ref{validUntrapFig} as a function of the interaction strength and the time. The area on the right of the black dashed line is forbidden because the quantity we plotted gets larger than one. 
The validity condition in the high-temperature regime is formally equivalent, apart from a factor $k_{\rm B}T/\hbar\omega_{\rm B}$ multiplying the left hand-side, inducing a constraint also on the temperature. 

\section{Zero-trap frequency limit}\label{LimitApp}
The results obtained in this manuscript regard an impurity embedded in a trapped BEC. 
Precisely we consider a harmonic confining potential, characterized by a frequency $\omega_{\rm B}$.  

A valid question, is, whether by taking the limit in which the BEC trapping frequency goes to zero, we recover the results presented in \cite{Lampo2017} for a homogeneous gas. We point out that the values of the position variance calculated in the two different situations do not match as $\omega_{\rm B}$ tends to zero.  
However, it is possible to note that this kind of pathology goes beyond our treatment since already occurs at the level of the Bogoliubov spectrum.
In fact our results rely on Eq.\ \eqref{BogSpec}, derived in \cite{Stringari1996, Ohberg1997}.
Here, we do not recover the traditional spectrum for a homogeneous gas, by sending $\omega_{\rm B}\rightarrow0$.

The impossibility to switch continuously from the inhomogeneous case to the homogeneous one, may also be understood in terms of the density of the bath states
\begin{equation}
\rho(\omega)=\sum_i\delta(\omega_i-\omega),
\end{equation}    
where $\omega$ is the frequency of the Bogoliubov modes in the continuous limit. 
By recalling Eq.\ \eqref{BogSpec} we get the expression of the density of states associated to an inhomogeneous gas:
\begin{equation}
\rho^{{\rm (Inh)}}=2\omega/\omega^2_{\rm B}.
\end{equation}
In a similar way we derive that for a homogeneous gas we have
\begin{equation}
\rho^{{\rm  (Hom)}}=\frac{V}{2\pi c},
\end{equation}
where $c$ is the speed of sound and $V$ the volume where is confined the homogeneous medium. 
The density of bath states shows two different expressions in the homogeneous and inhomogeneous case (it is interesting to note that their ratio is proportional to that between the corresponding SDs, \textit{i.e.} $\rho^{{\rm (Inh)}}/\rho^{{\rm (Hom)}}\sim\omega$). 
Hence, we approach a very similar situation to that of 2D ideal gas, where the different form of the density of states arising in the presence of a trap does not exhibit a continuous crossover to the case without trap~\cite{Petrov2004} (e.g. in the trapped case there is actually condensation while in the homegeneous case not). 

In order to match the physics of the homogeneous case in the zero-trap frequency we could properly study the scaling of the several quantities involved in the physics of the system. 
Precisely one may aim to get the linear branch of the Bogoliubov spectrum of the homogeneous gas by taking in Eq.\ \eqref{BogSpec} both the limit $\omega_{\rm B}\rightarrow0$ and $j\rightarrow\infty$, keeping constant their product  $\omega_{\rm B}j=ck$. 
Nevertheless, although one reproduces the same spectrum, such a procedure does not work for the relative eigenstates\ \eqref{fMin1D}, and thus for the interaction Hamiltonian\ \eqref{IntHamMiddle}. 
From the formal point of view this is due to the difficulty of obtaining plane waves from the Legendre polynomials in the zero-trap frequency. 
In fact the same problem emerges already for the physics of single particle: once one solves the Shr\"odinger equation for the harmonic oscillator, it is not possible to recover the eigenstates of the free particle (plane waves) just by sending the frequency to zero. 

Finally, the possibility of performing the zero-trap frequency limit is also affected by the limits of the Thomas-Fermi regime, on which our analysis is based. 
First of all, the Thomas-Fermi density profile\ \eqref{TFDensProf} constitutes the solution of the Gross-Pitaevskii in the limit in which we drop out the kinetic term. 
In this context the zero-trap frequency limit is equivalent to sending to zero the potential energy, resulting in a system with zero energy, which is meaningless. 
Note in fact that the density\ \eqref{TFDensProf}, as well as the spectrum\ \eqref{BogSpec}, goes to zero in this limit, namely we are turning off the bath. 

Furthermore, the Thomas-Fermi approximation holds when the physics of the gas is ruled by the trapping confinement rather than the interparticles interaction. 
In the zero-trap frequency limit we have a situation strongly governed by the interaction and so the Thomas-Fermi approximation fails .
According to this, it is possible to evaluate the threshold trap frequency below which our analysis is no longer faithful.
This task has been realized in \cite{Petrov2004} where the parameter
\begin{equation}
\alpha=\frac{m_{\rm B}g_{\rm B}a_{\rm HO}}{\hbar^2},\quad a_{\rm HO}=\frac{\hbar}{m_{\rm B}\omega_{\rm B}},
\end{equation}
was introduced.
Thomas-Fermi approximation is ensured if the condition 
\begin{equation}\label{TFValCond}
N\gg\alpha^2
\end{equation}
is fulfilled, otherwise the medium passes to the strong-coupling regime (see Fig. 5 in \cite{Petrov2004}). 
In this way one may infer the trap frequency threshold. 
We see, however, that in the limit in which such a frequency goes to zero the condition in \eqref{TFValCond} fails. 

\bibliography{QBM_in_a_BEC}
\end{document}